# A Detailed Investigation of the Onion Structure of Exchanged Coupled Magnetic $Fe_{3-\delta}O_4$@$CoFe_2O_4$@$Fe_{3-\delta}O_4$ Nanoparticles

Kevin Sartori,[1,2,3] Anamaria Musat,[1] Fadi Choueikani,[2] Jean Marc Grenèche,[4] Simon Hettler,[5,6] Peter Bencok,[7] Sylvie Begin-Colin,[1] Paul Steadman,[7] Raul Arenal,[5,6,8] Benoit P. Pichon[*,1,9]

[1] Université de Strasbourg, CNRS, Institut de Physique et Chimie des Matériaux de Strasbourg, UMR 7504, F-67000 Strasbourg, France
[2] Synchrotron SOLEIL, L'Orme des Merisiers, Saint Aubin – BP48, 91192 Gif-sur-Yvette, France
[3] Laboratoire Léon Brillouin, UMR12 CEA-CNRS, F-91191 Gif-sur-Yvette, France
[4] Institut des Molécules et Matériaux du Mans, IMMM, UMR CNRS-6283 Université du Maine, avenue Olivier Messiaen, 72085 Le Mans Cedex 9, France
[5] Instituto de Nanociencia y Materiales de Aragon (INMA), CSIC-Universidad de Zaragoza, Calle Pedro Cerbuna, 50009 Zaragoza, Spain
[6] Laboratorio de Microscopias Avanzadas (LMA), Universidad de Zaragoza, Calle Mariano Esquillor, 50018 Zaragoza, Spain
[7] Diamond Light Source, Didcot OX11 0DE, UK
[8] Fundacion ARAID, 50018 Zaragoza, Spain
[9] Institut Universitaire de France, 1 rue Descartes, 75231 Paris Cedex 05, France

Corresponding Author

*E-mail: Benoit.Pichon@unistra.fr

**Abstract**

Nanoparticles which combine several magnetic phases offer wide perspectives for cutting edge applications because of the high modularity of their magnetic properties. Besides the addition of the magnetic characteristics intrinsic to each phase, the interface that results from core-shell and, further, from onion structures leads to synergistic properties such as magnetic exchange coupling. Such a phenomenon is of high interest to overcome the superparamagnetic limit of iron oxide nanoparticles which hampers potential applications such as data storage or sensors. In this manuscript, we report on the design of nanoparticles with an onion-like structure which have been scarcely reported yet. These nanoparticles consist in a $Fe_{3-\delta}O_4$ core covered by a first shell of $CoFe_2O_4$ and a second shell of $Fe_{3-\delta}O_4$, *e.g.* a $Fe_{3-\delta}O_4@CoFe_2O_4@Fe_{3-\delta}O_4$ onion-like structure. They were synthesized by a multi-step seed mediated growth approach which consists to perform three successive thermal decomposition of a metal complexes in a high boiling point solvent (about 300 °C). Although TEM micrographs clearly show the growth of each shell from the iron oxide core, core sizes and shell thicknesses markedly differ from what is suggested by the size increase. We investigated very precisely the structure of nanoparticles in performing high resolution (scanning) TEM imaging and geometrical phase analysis (GPA). The chemical composition and spatial distribution of atoms were studied by electron energy loss spectroscopy (EELS) mapping and spectroscopy. The chemical environment and oxidation state of cations were investigated by Mössbauer spectrometry, soft X-ray absorption spectroscopy (XAS) and X-ray magnetic circular dichroism (XMCD). The combination of these techniques allowed us to estimate the increase of $Fe^{2+}$ content in the iron oxide core of the core@shell structure and the increase of the cobalt ferrite shell thickness in the core@shell@shell one, while the iron oxide shell appears to be much thinner than expected. Thus, the modification of the chemical composition as well as the size of the $Fe_{3-\delta}O_4$ core and the thickness of the cobalt ferrite shell have a high impact on the magnetic properties. Furthermore, the growth of the iron oxide shell also markedly modifies the magnetic properties of the core-shell nanoparticles, thus demonstrating the high potential of onion-like nanoparticles for tuning accurately the magnetic properties of nanoparticles according to the desired applications.

**Introduction**

Bimagnetic nanoparticles open huge perspectives toward potential applications in fields such as biomedicine, sensors, or data storage because of the high modulation of their magnetic properties.[1] It is very well established that, at the nanoscale, the surface contribution predominates on the volume contribution.[2] Therefore, slight modifications of the size and the shape significantly influence the magnetic properties. Besides controlling the size and the shape of nanoparticles, the design of multicomponent nanoparticles allows the intrinsic magnetic properties of different phases to be combined. In addition, core@shell structures which result in large interfaces at the nanoscale usually favor synergistically enhanced magnetic properties such as effective magnetic anisotropy energy. This approach represents a high potential to reduce the amount of rare earths used to produce permanent magnets and classified by the European Union as critical raw materials.[3,4,5] Metal oxide nanoparticles such as iron oxide ($Fe_{3-\delta}O_4$, magnetite / maghemite)[6] - which is cheap, nontoxic and abundant – represent an interesting alternative. Although the magnetic anisotropy energy is not high enough to produce permanent magnets at room temperature – iron oxide nanoparticles are superparamagnetic-, it can be markedly enhanced by coating nanoparticles with a harder magnetic metal oxide.[7–12,13] The design of core-shell nanoparticles gives rise to an interesting interfacial magnetic properties which consists in the pinning of soft spins of the iron oxide core by the harder spin of the harder shell, e.g. the so-called exchange bias coupling.[14] This phenomenon is of great interest to push the superparamagnetic limit over room temperature.[15] $Fe_{3-\delta}O_4$ is usually combined to antiferromagnetic phases such as CoO that have a magnetic anisotropy constant which is two order of magnitude higher than the one of $Fe_{3-\delta}O_4$.[16,16–20,21] Nevertheless, exchange bias coupling only happens below the Néel temperature ($T_N$ = 290 K for CoO). Indeed, above the Néel temperature, the antiferromagnetic order vanishes and loses its ability to pin the spins of the soft phase.

In contrast, ferrite ($MFe_2O_4$) phases open interesting perspectives because of their common spinel structure and close lattice parameters, while their magnetic hardness and softness markedly depend on the transition metal M.[22] Such ferrimagnetic (FiM) materials display Curie temperatures that are usually much higher than room temperature ($T_C$ = 790 K for $CoFe_2O_4$), thus avoiding the Néel temperature limitation of antifferromagnetic phases. Exchange coupled nanoparticles which combine several ferrite phases into a core-shell structure showed remarkably tunable magnetic properties such as enhanced magnetic anisotropy and magnetization saturation.[23,24–30] The exceptional properties of Ferrite@Ferrite nanoparticles result from much more complex structure than the ideal picture of a well-define interface in core-shell structure. Although each crystal phase is selected because of the low lattice mismatch, defects may occur at the interface because of the shape of the nanoparticles.[31] Indeed, an isotropic shape which is close to sphere induces a high curvature radius and facets with different surface energies which result in the complex growth of the shell component at the core surface. Considering defects at the surface of the nanoparticle which result from the break of symmetry, diffusion of cations leading to an interfacial composition gradient between the core and the shell has been regularly reported.[8,32–34] The growth of the shell usually occurring at high temperatures (200 – 300 °C) results from a synthesis mechanism which does not systematically consist in a simple seed mediated growth process. Indeed, it usually consists in the partial solubilization of the seeds which is usually followed by a recrystallization of the monomers issued from the Ostwald ripening process and with those that remain from the decomposition of the reactant.[34,35] All these features significantly alter the expected ideal chemical composition of the core-shell. Therefore, the magnetic properties deviate from those initially expected and are difficult to anticipate.

Non-hydrolytic synthesis techniques have been demonstrated to be particularly effective to design core-shell nanoparticles with a very high control of their structure. Hence, the effect of the core size and of the shell thickness on their magnetic properties can be systematically studied.[8,19,20,24,29] The new challenge is to design nanoparticles with more complex structure in order to precisely tune their magnetic properties. In this aim, onion-like structures, which consist of a core covered by several shells, have recently driven a tremendous interest, although they have been scarcely reported so far.[36–38] Although the synthesis of such complex nanoparticles is not trivial, the fine understanding of the onion structure is necessary to rationalize the study of their magnetic properties. Recently, we reported on the synthesis of nanoparticles which consist in a $Fe_{3-\delta}O_4@CoO@Fe_{3-\delta}O_4$ onion-like structure.[39] Both soft/hard and hard/soft interfaces resulted in blocked magnetization at room temperature, although these nanoparticles are mostly composed of iron oxide with a size below 16 nm. Such enhanced magnetic properties account from a more complex structure that resulted from the partial replacement of CoO by cobalt ferrite at both interfaces during the synthesis steps.[40]

In this context, we have designed new multi-component nanoparticles in order to rationally investigate their structure by combining advanced characterization techniques. Hence, we report here on onion-like magnetic nanoparticles which consist of an iron oxide core combined with a first shell of cobalt ferrite and a second shell of iron oxide, i.e. a $Fe_{3-\delta}O_4@CoFe_2O_4@Fe_{3-\delta}O_4$ structure. We have deeply studied the structure in order to bring a better understanding on its relationship with the magnetic properties. The onion-like structure of nanoparticles was systematically compared to those of seeds – iron oxide and core@shell nanoparticles – by means of highly complementary and advanced characterization techniques. Spatially-resolved energy electron loss spectroscopy (EELS) mapping and spectrum analysis, Mössbauer spectrometry, X-ray absorption spectroscopy (XAS) and X-ray magnetic circular dichroism (XMCD) were used in order to accurately characterize the chemical composition and the cationic distribution. The combination of such advanced techniques allowed us to show that the core size and the shell thicknesses markedly differ from what is suggested by size variations observed on TEM micrographs. Finally, the magnetic properties were correlated to the structure of the nanoparticle, in order to evaluate the effect of the modification of the iron oxide core size, the cobalt ferrite shell thickness and the growth of the second iron oxide shell.

**Experimental section**

**Metal (Fe or Co) stearate synthesis.** Iron stearate was synthesized according to our previous work[41] while the synthesis of cobalt stearate was adapted. In a 1 L two-necked round bottom flask, 9.8 g (32 mmol) of sodium stearate (98.8 %, TCI) were poured and 320 mL of distilled water were added. The mixture was heated to reflux under magnetic stirring until all the stearate was dissolved. Afterwards, 3.80 g (16 mmol) of iron (II) chloride tetrahydrated (or 3.16 g (16 mmol) of cobalt (II) chloride hexahydrated) dissolved in 160 mL of distilled water were poured in the round bottom flask. The mixture was heated to reflux and kept to this temperature for 15 minutes under magnetic stirring before cooling down to room temperature. The colored precipitate was collected by centrifugation (15 000 rpm, 5 min) and washed by filtration with a Buchner funnel. Finally, the powder was dried in an oven at 65 °C for 15 hours.

**Nanoparticle synthesis.** $Fe_{3-\square}O_4@CoFe_2O_4@Fe_{3-\square}O_4$ nanoparticles were synthesized using a three steps thermal decomposition method in a similar way that we reported recently.[39] First, iron oxide nanoparticles were synthesized according to our previous work.[42] A two-necked round bottom flask was filled with 1.38 g (2.22 mmol) of iron (II) stearate, 1.254 g (4.44 mmol) of oleic acid (99% Alfa Aesar) and 20 mL of ether dioctyl ($B_P$ = 290 °C, 97 % Fluka). The brownish mixture was heated at 100 °C under a magnetic stir for 30 min in order to remove water residues and to homogenize the solution. The magnetic stirrer was then removed and the flask was connected to a reflux condenser before heating the solution to reflux for 2 h with a heating ramp of 5°C/min. At the end, the mixture was allowed to cool down to 100 °C and 4 mL of the solution were removed and washed to serve as a reference (sample C).

Secondly, 0.29 g (0.46 mmol) of cobalt (II) stearate, 0.791 g (2.8 mmol) of oleic acid and 32 mL of 1-octadecene were added to the reaction medium. The mixture was heated to 100 °C for 30 min under magnetic stirring to remove water residues and to homogenize the solution. After removal of the magnetic stirrer, 0.585 g (0.94 mmol) of iron (II) stearate was added. The flask was then connected to a reflux condenser in order to heat the solution at reflux for another 2 h with a heating ramp of 1 °C/min. After cooling down to room temperature, the nanoparticles were precipitated by the addition of acetone in order to wash them by centrifugation with a mixture of chloroform: acetone (1 : 5). The final nanoparticles (sample CS) were stored in chloroform.

Thirdly, half of the volume of the washed CS suspension was poured in a two-necked round bottom flask with 0.548 g (0.88 mmol) of iron (II) stearate, 0.497 g (1.76 mmol) of oleic acid and 20 mL of ether dioctyl. The mixture was then heated to 100 °C under magnetic stirring for 30 min. As mentioned above, after removing the magnetic stirrer, the mixture was heated to reflux for 2 h with a heating ramp of 1 °C/min. After cooling down, the nanoparticles were collected and washed in the same way as for CS nanoparticles. The final nanoparticles (sample CSS) were stored as a colloidal suspension in chloroform.

**Transmission electron microscopy (TEM)** was performed by using a JEOL 2100 LaB6 with a 0.2nm point-to-point resolution and a 200 KV acceleration voltage. EDX was performed with a JEOL Si(Li) detector. The average size of the nanoparticles was calculated in measuring at least 300 nanoparticles from TEM micrographs by using the Image J software. The average shell thickness was calculated as the half of the difference between the size of the nanoparticles before and after the thermal decomposition step. The size distribution was fitted by a log-normal function.

**Scanning transmission electron microscopy (STEM)** experiments were carried out using a probe aberration corrected Titan (Thermo Fisher Scientific) equipped with a high-brightness field emission gun. While the electron gun was operated at 300 keV for acquisition of high-angle annular dark field (HAADF, acceptance angle 47.9 mrad) STEM images to obtain maximum spatial resolution (convergence angle 25 mrad), the high energy was lowered to 80 keV for electron energy-loss spectroscopy (EELS) to minimize beam damage, to increase the EELS signal and to improve energy resolution (~1 eV). The Gatan imaging filter (GIF, Gatan Inc) was operated at a dispersion of 0.2 eV /px in order to simultaneously analyze O$K$, Fe$L$ and Co$L$ edges with a collection angle of 119 mrad. EELS spectra and spectrum images (SI) were treated using a custom Matlab software including principal component analysis (PCA) for noise reduction. Quantification was done using power-law background subtraction and an integration width of 40 eV for the C and 30 eV for the CS/CSS nanoparticles. The sample preparation was done by drop casting 2 µL of the nanoparticle suspension on Holey-C grids followed by 14 s of plasma cleaning.

**X-ray diffraction (XRD)** was performed using a Bruker D8 Advance equipped with a monochromatic copper radiation (K$\alpha$ = 0.154056 nm) and a Sol-X detector in the 20– 80° 2θ range with a scan step of 0.02°. High purity silicon powder (a = 0.543082 nm) was systematically used as an internal standard. Crystal sizes were calculated by the Scherrer's equation and cell parameters by the Debye's law.

**Fourier transform infra-red** (FT-IR) spectroscopy was performed using a Perkin Elmer Spectrum spectrometer in the energy range 4000–400 cm$^{-1}$ on samples diluted in KBr pellets.

**Granulometry** measurements were performed using a nano-sizer Malvern (nano ZS) zetasizer at a scattering angle of 173°. A measure corresponds to the average of 7 runs of 30 seconds.

**Themogravimetry** analyses (TGA) were performed using a SDTQ600 from TA instrument. Measurements were performed on dried powders under air in the temperature range of 20 to 600 °C at a heating rate of 5 °C/min.

**X-ray absorption.** XAS and XMCD spectra were recorded at the $L_{2,3}$ edges of Fe and Co, on the DEIMOS beamline at SOLEIL (Saclay, France)[43] and on I10 (BLADE) beamline at Diamond Light Source (Oxford, United Kingdom). All spectra were recorded at 4.2 K under UHV conditions (10$^{-10}$ mbar) and using total electron yield (TEY) recording mode. The measurement protocol was previously detailed by Daffé et al.[44] An external parallel magnetic field H$^+$ (respectively antiparallel H$^-$) was applied on the sample while a σ$_+$ polarized (σ$_-$ polarized respectively) perpendicular beam was directed on the sample. Isotropic XAS signals were obtained by taking the mean of the σ$_+$+σ$_-$ sum where σ$_+$ = [σ$_L$(H$^+$)+ σ$_R$(H$^-$)]/2 and σ$_-$ = [σ$_L$(H$^-$)+ σ$_R$(H$^+$)]/2 with σ$_L$ and σ$_R$ the absorption cross section measured respectively with left and right circularly polarized X-rays. XMCD spectra were obtained by taking the σ$_+$-σ$_-$ dichroic signal with a ± 6.5 T applied magnetic field.

At DEIMOS beamline, the circularly polarized X-rays are provided by an Apple-II HU-52 undulator for both XAS and XMCD measurements while EMPHU65 with a polarization switching rate of 10 Hz was used to record hysteresis cycle at fixed energy.[43]. Measurements were performed between 700 and 740 eV at the iron edge and between 770 and 800 eV at the cobalt edge with a resolution of 100 meV and a beam size of 800*800 µm. Both XMCD and isotropic XAS signals presented here are normalized by dividing the raw signal by the edge jump of the isotropic XAS.

At BLADE beamline, the circularly polarized X-rays were provided by a helical undulator with 48 cm periodicity. Monochromatic X-rays in the soft X-ray range (400-1600 eV) were provided with a plane grating monochromator[45] beamline giving an energy resolution of 100meV and a beam size of 100*100µm$^2$ (root mean square) at the sample position. Sample cooling and applied field were supplied with an Oxford Instruments cryomagnet.

The samples consist of drop casted suspension of nanoparticles in chloroform onto a silicon substrate. The substrates were then affixed on a sample holder.

**Mössbauer spectrometry.** $^{57}$Fe Mössbauer spectra were performed at 77 K using a conventional constant acceleration transmission spectrometer with a $^{57}$Co(Rh) source and a bath cryostat. The samples consist of 5 mg Fe/cm$^2$ powder concentrated in a small surface due to the rather low quantities. The spectra were fitted by means of the MOSFIT program[46] involving asymmetrical lines and lines with Lorentzian profiles, and an α-Fe foil was used as the calibration sample. The values of isomer shift are quoted relative to that of α-Fe at 300 K.

**SQUID magnetometry.** Magnetic measurements were performed in using a Superconducting Quantum Interference Device (SQUID) magnetometer (Quantum Design MPMS-XL 5). Temperature dependent zero-field cooled (ZFC) and field cooled (FC) magnetization curves were recorded as follows: the sample was introduced in the SQUID at room temperature and cooled down to 5 K with no applied magnetic field and after applying a careful degaussing procedure. Then, a magnetic field of 7.5 mT was applied, and the ZFC magnetization curve was recorded upon heating from 5 to 400 K. The sample was then cooled down to 5 K under the same applied field, and the FC magnetization curve was recorded upon heating from 5 to 400 K. Magnetization curves as a function of a magnetic field (M(H) curves) applied in the plane of the substrate were measured at 5 and 400 K. The sample was also introduced in the SQUID at high temperature and cooled down to 5 K with no applied magnetic field (ZFC curve) and after applying a subsequent degaussing procedure. The magnetization was then measured at constant temperature by sweeping the magnetic field from +7 T to −7 T, and then from −7 T to +7 T. To evidence exchange bias effect, FC M(H) curves were further recorded after heating up at 400 K and cooling down to 5 K under a magnetic field of 7 T. The FC hysteresis loop was then measured by applying the same field sweep as for the ZFC curve. The coercive field ($H_C$) and the $M_R/M_S$ ratio were measured from ZFC M(H) curves. The exchange bias field ($H_E$) was measured from FC M(H) curves. Magnetization saturation ($M_S$) was measured from hysteresis recorded at 5 K.

## Results and discussion

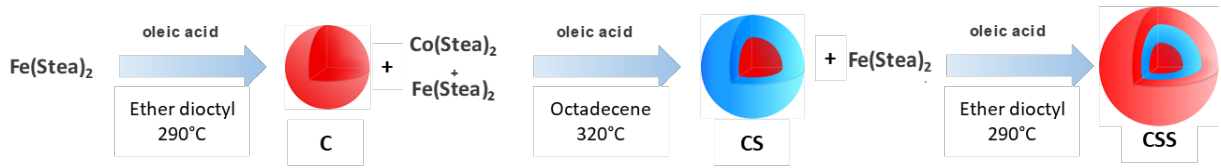

**Figure 1.** Schematic illustration of the synthesis pathway of onion-like nanoparticles (CSS) using a three-step thermal decomposition method.

Core@shell@shell nanoparticles were synthesized by performing successively the thermal decomposition of a metal stearate three times (Figure 1). First, the thermal decomposition of iron (II) stearate (FeSt$_2$) in ether dioctyl (B$_P$ = 290 °C) was performed in presence of oleic acid in order to synthesize Fe$_3$O$_4$ nanoparticles (C). Second, cobalt (II) stearate (CoSt$_2$) and FeSt$_2$ (molar ratio 1:2) were decomposed together in octadecene (B$_P$ = 320 °C), in order to grow a cobalt ferrite (CoFe$_2$O$_4$) shell at the surface of pristine Fe$_3$O$_4$ nanoparticles. The aim was to synthesize core-shell Fe$_3$O$_4$@CoFe$_2$O$_4$ nanoparticles (CS). Finally, FeSt$_2$ was again decomposed in ether dioctyl in the presence of CS nanoparticles with the aim to grow a second Fe$_3$O$_4$ shell, i.e. to synthesize Fe$_3$O$_4$@CoFe$_2$O$_4$@Fe$_3$O$_4$ (CSS) nanoparticles.

TEM micrographs (Figure 2) show that C nanoparticles display a homogeneous shape close to sphere and a narrow size distribution centered at 8.0 ± 0.9 nm. The nanoparticle size increases from CS (10.0 ± 1.5 nm) to CSS (13.1 ± 2.2 nm) corresponding to average shell thicknesses of 1.0 nm and 1.6 nm, respectively. The broadening of size distribution and the deviation of shape from sphere to CS and CSS is ascribed to the inhomogeneous growth of both shells. Indeed, the nanoparticle surface consists in facets which feature different surface energies according to the corresponding *hkl* planes.

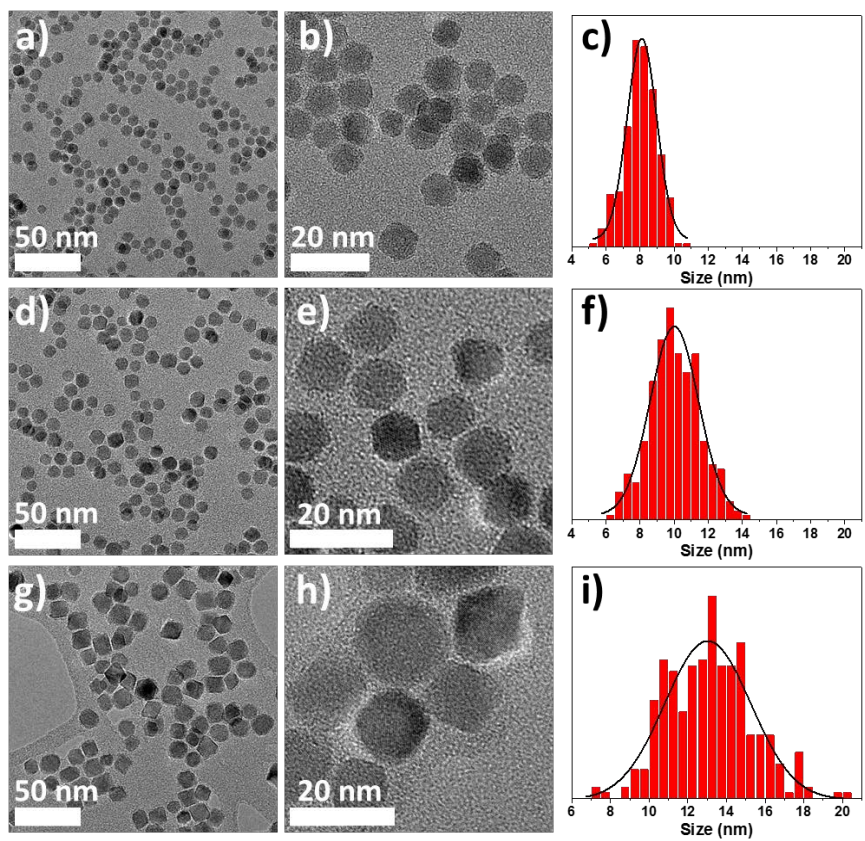

**Figure 2.** Conventional TEM micrographs of a), b) C nanoparticles; d), e) CS nanoparticles and g), h) CSS nanoparticles with (c), f), i)) their corresponding size distributions.

**Table 1.** Structural characteristics of nanoparticles. Mean core sizes and shell thicknesses were calculated from TEM micrographs. Cell parameters and crystal sizes were calculated from XRD patterns.

|  | C | CS | CSS |
|---|---|---|---|
| **Size (nm)** | 8.0 ± 0.9 | 10.0 ± 1.5 | 13.1 ± 2.2 |
| **Size variation (nm)** | - | 2.0 | 3.1 |
| **Fe : Co at. Ratio by EDX** | - | 86 : 14 | 94 : 6 |
| **Hydrodynamic diameter (nm)** | 8.7 | 13.5 | 18.2 |
| **Cell parameter (Å)** | 8.37 ± 0.01 | 8.41 ± 0.01 | 8.41 ± 0.01 |
| **Crystal size (nm)** | 7.4 ± 0.5 | 10.1 ± 0.5 | 12.0 ± 0.5 |

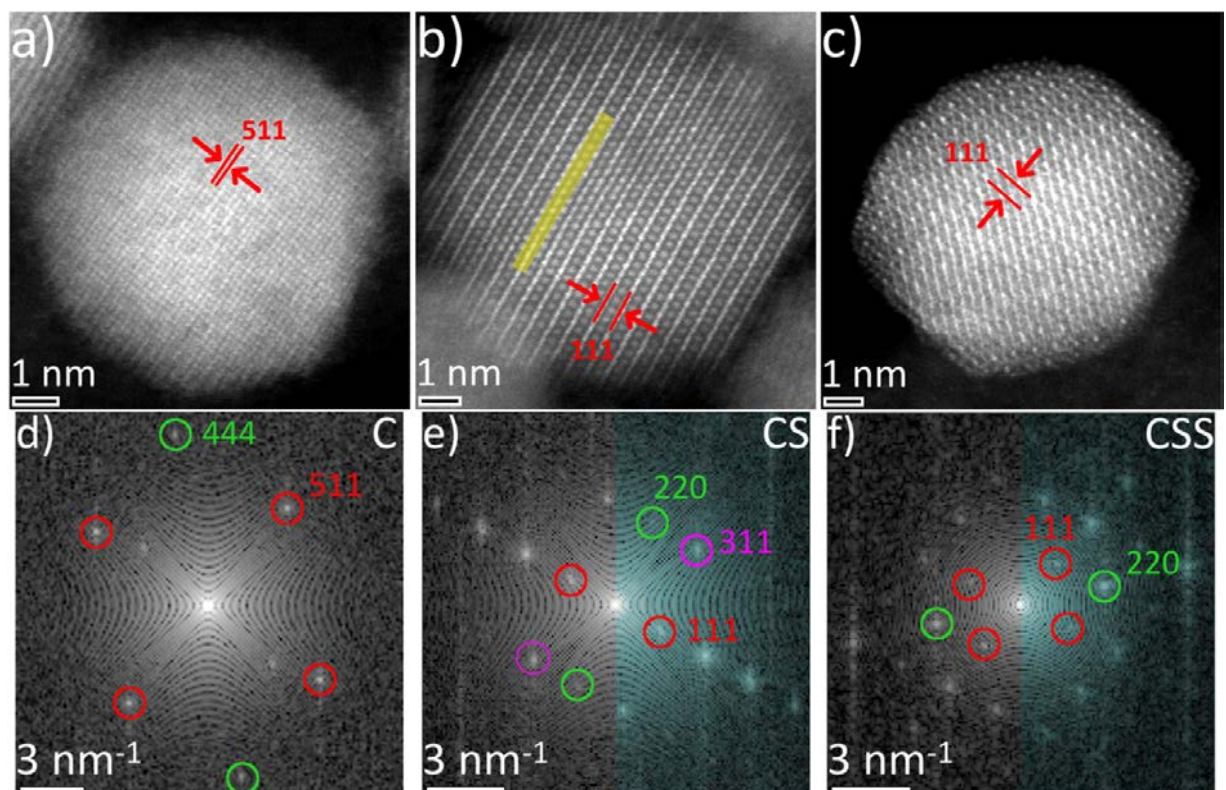

**Figure 3.** STEM-HAADF micrographs of a) C, b) CS and c) CSS nanoparticles showing their microstructures. Inter-reticular distances are highlighted by double red arrows. A stacking defect may be observed by the change of the atomic column contrast along the highlighted area in yellow, also visible in neighboring lines. (d)-f)) FFT from the C, CS and CSS nanoparticles obtained from the core region of the nanoparticle using a circular smoothed mask. FFT from the shell region are overlaid in green color in the right half for CS and CSS nanoparticles. Colored circles evidence the related *hkl* reflections attributed to magnetite (JCPDS card n° 19-062) and to cobalt ferrite (JCPDS card n° 00-022-1086). Colors refer to *hkl* plan families.

STEM-HAADF micrographs of C, CS and CSS nanoparticles display straight and continuous lattice fringes, evidencing the single crystal-like structure of the nanoparticles resulting from the successive epitaxial growth of the different shells (Figures 3a-c). Minor crystal defects were observed in a few nanoparticles, e.g. in the CS nanoparticle (Figure 3b). These defects are recognizable by the changing contrast of the atomic columns from dots to a line along the highlighted area. As the pattern remains undisturbed, the defects are attributed to stacking defects. Figure S1 shows additional STEM micrographs of the nanoparticles suggesting that minor defects were already present in few of the C nanoparticles, although an induction by electron beam damage cannot be excluded. The inter-reticular distances between two fringes were attributed to the spinel ferrite phase (including magnetite, maghemite and cobalt ferrite) for each sample. These results are supported by FFT calculated from STEM-HAADF micrographs that show spots corresponding to (*hkl)* directions of the spinel phase (Figures 3d-f). A comparison of FFT calculated from core and shell regions revealed perfect overlap of the spots, agreeing with good epitaxial relationships.

Energy dispersive X-ray spectrometry (EDX) showed that CS nanoparticles consist of Fe (86 at. %) and Co (14 at. %) which agree with values calculated for a $Fe_{3-\square}O_4$ core of 8.0 nm (Fe: 84 at. %) and a $CoFe_2O_4$ shell thickness of 1.0 nm (Co: 16 at. %), as measured from TEM micrographs. CSS nanoparticles display the increase of Fe (94 at. %) vs. Co (6 at. %) ratio, in agreement with the size variation measured from TEM micrographs, which corresponds to the growth of iron oxide at the surface of CS (Fe : 93 at. % and Co : 7 at. %). Nevertheless, EDX does not give any information on the spatial arrangement of Fe and Co within the nanoparticle volume. Therefore, we performed complementary measurements using advanced characterization techniques.

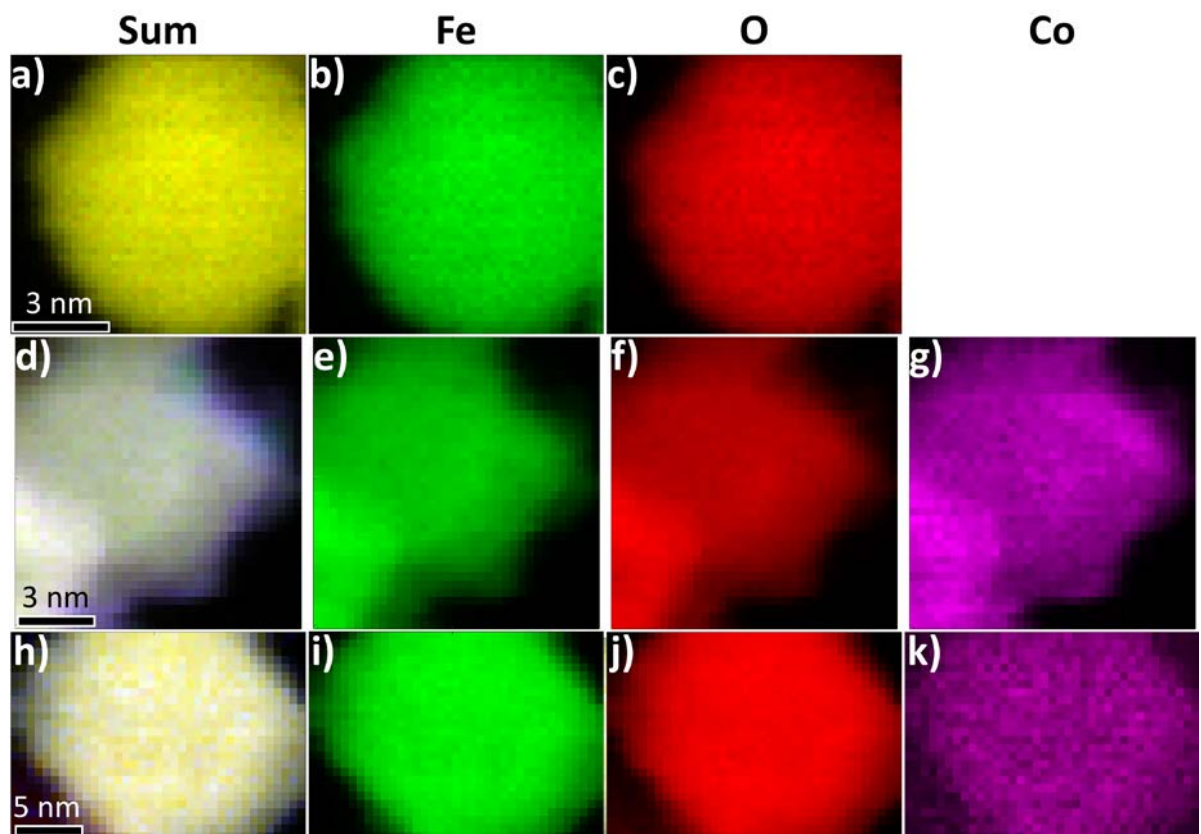

**Figure 4.** Elemental mapping performed by EELS-SI on isolated nanoparticles (a)-c)) C, (d)-g)) CS, (h)-k)) CSS with a), d), h) the sum of the composite (Fe in green + Oxygen in red + Cobalt in blue), b), e), i) Fe-edge, c), f), j) O-edge and g), k) Co-edge, which is displayed in magenta to improve visibility compared to blue.

The spatial distribution of Fe, O and Co atoms was investigated by performing elemental mapping (~ 0.2 nm resolution) with electron energy loss spectroscopy spectrum-imaging (EELS-SI) of the Fe L-edge (green), Co L-edge (magenta, blue in composite) and O K-edge (red) (Figure 4). EELS-SI micrographs of C nanoparticles evidence the homogeneous atomic distribution of Fe and O atoms all across the nanoparticle which agrees with an iron oxide structure. In the case of CS nanoparticles, Fe, O and Co spatial distributions also overlap. A slight increase of the Co content at the edges of the nanoparticle (blue border of NP in Figure 4d) was observed, which agrees with the expected $Fe_{3-\square}O_4@CoFe_2O_4$ core@shell structure. Finally, CSS nanoparticles display a homogeneous spatial distribution of Fe, Co and O atoms all across the nanoparticle. The Fe/Co atomic ratio tends to increase in comparison to CS nanoparticles, in agreement with the expected $Fe_{3-\square}O_4@CoFe_2O_4@Fe_{3-\square}O_4$ structure and in line with EDX results. Line profiles of the composition across CS and CSS confirmed that the Co content increases at the edges of both nanoparticles (Figures S2 and S3). Therefore, the second shell of iron oxide is much thinner than the size variation between CS and CSS (3.1 nm).

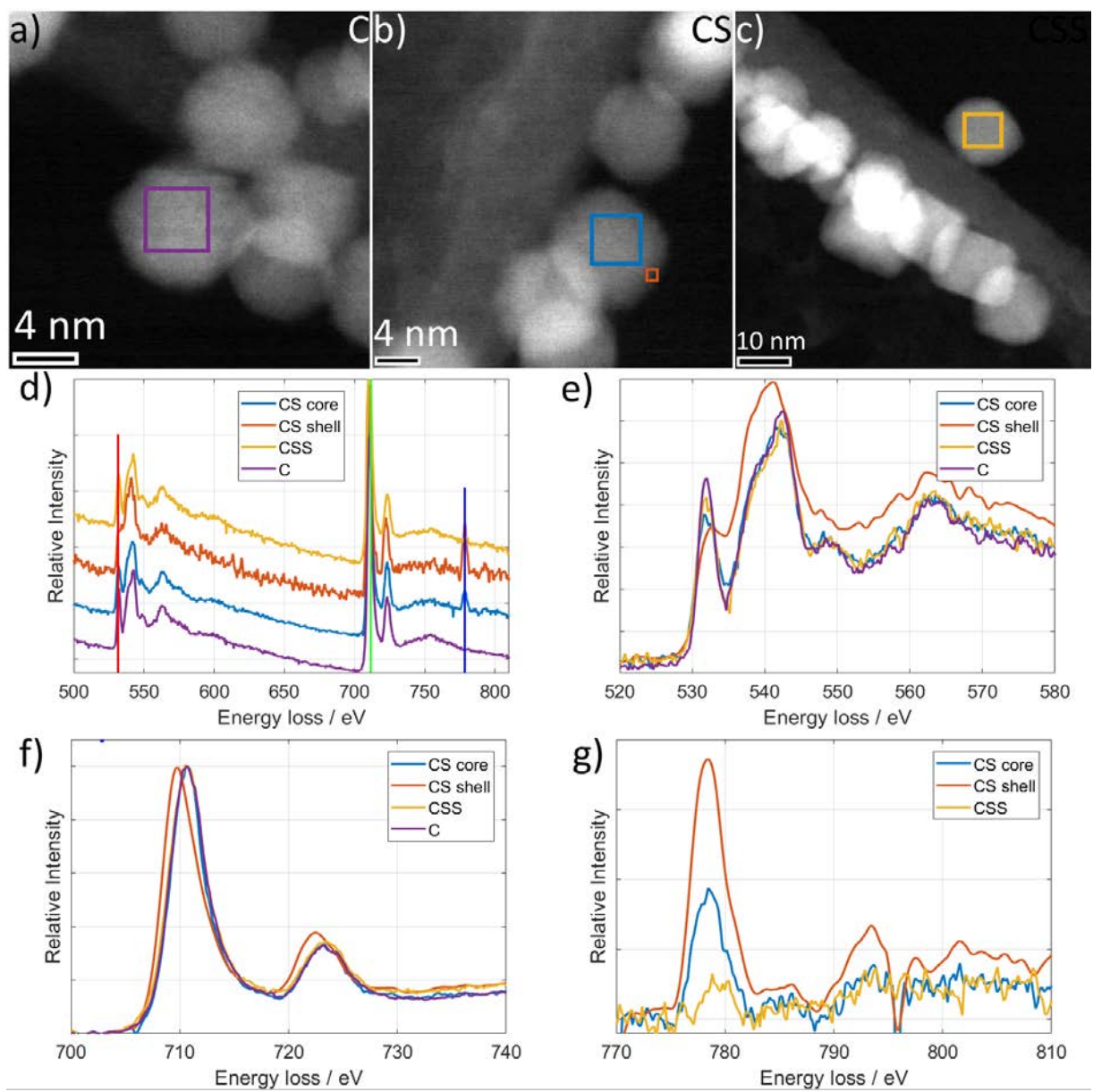

**Figure 5.** Dark field images of a) C, b) CS and c) CSS nanoparticles. d) Exemplary EELS spectra obtained from positions marked in (a)-c)) showing the O-K edge (532 eV, red vertical line), Fe-L (708 eV, green vertical line) edges as well as the Co-L edge (779 eV, blue vertical line) in case of CS and CSS nanoparticles. The spectra are vertically displaced to improve visibility. e)-g) Comparison of the background-subtracted EELS spectra of the e) O-K, f) Fe-L and g) Co-*L* edges reveal fine structure changes induced by the presence of Co in comparison to the C spectra. The Fe-$L_3$ peak shifts to lower energies in case of high Co percentages (CS shell). All spectra are normalized to the Fe-$L_3$ peak.

To further investigate the chemical composition of C, CS and CSS nanoparticles, EELS spectra were recorded at precise positions within the EELS-SI data by averaging over the corresponding area, excluding the shell (Figure 5). Spectra recorded for C (magenta) do not vary within the nanoparticle and corresponds to a homogeneous chemical composition of $Fe_{3-\square}O_4$ (Figure S5).[47] In contrast, spectra of CS clearly show a higher Co (779 eV) / Fe (708 eV) intensity ratio at the edge than in the center of the nanoparticle (red and blue line in Figure 5d), which agrees with a $Fe_{3-d}O_4@CoFe_2O_4$ core@shell structure. In contrast, spectra recorded for CSS (yellow) nanoparticles show that O*K* and Fe*L* signals are closely related to that of the C spectrum and displays a weaker signal at the Co-L edge than CS. The Fe/Co ratio is thus increased compared to CS nanoparticles which is coherent with EDX results. A comparison of spectra obtained from different CS and CSS nanoparticles of the same batch reveals the reproducibility and homogeneity within the batches (Figures S6 and S7).

The background-subtracted oxygen signal of the same spectra is shown in Figure 5e and reveals a clear dependence on the different cation environments. While the appearance of the broad peak around 540 eV does not differ much between the spectra, the intensity of the sharp peak at 532 eV and the intensity of the following valley at 534 eV strongly vary with the Co content. The spectrum from the C nanoparticle displays a similar shape to the spectrum of magnetite $Fe_3O_4$ (Figure S5),[47–49] exhibiting a sharp and intense peak at 532 eV followed by a deep valley at 534 eV. With increasing the presence of Co in the material, the peak intensity at 532 eV decreases while the intensity in the valley at 534 eV increases. In the spectrum obtained from the shell of the CS nanoparticle, i.e. the area with the highest Co content, the intensities at 530 eV and 533 eV almost level out. This behavior is consistent with the one observed in CoO spectra.[50] The comparison of the local intensity ratio between the pre-peak and the following valley allows to map the chemical composition (Figure S8). While the O-K edge is highly sensitive to the Co content in the crystal, an effect on the Fe-L edge is only visible at considerable Co contents leading to a shift of the peak position to lower energies as it is observed in the shell of the CS nanoparticle (red and blue lines in Figure 5f). A fit of the Fe-L peak allows to determine the exact peak position and to map the chemical composition within the CS nanoparticles (Figure S8). Unfortunately, a fine edge analysis of the Co peak is not possible because of the low intensity in the Co peak (Figure 5g).

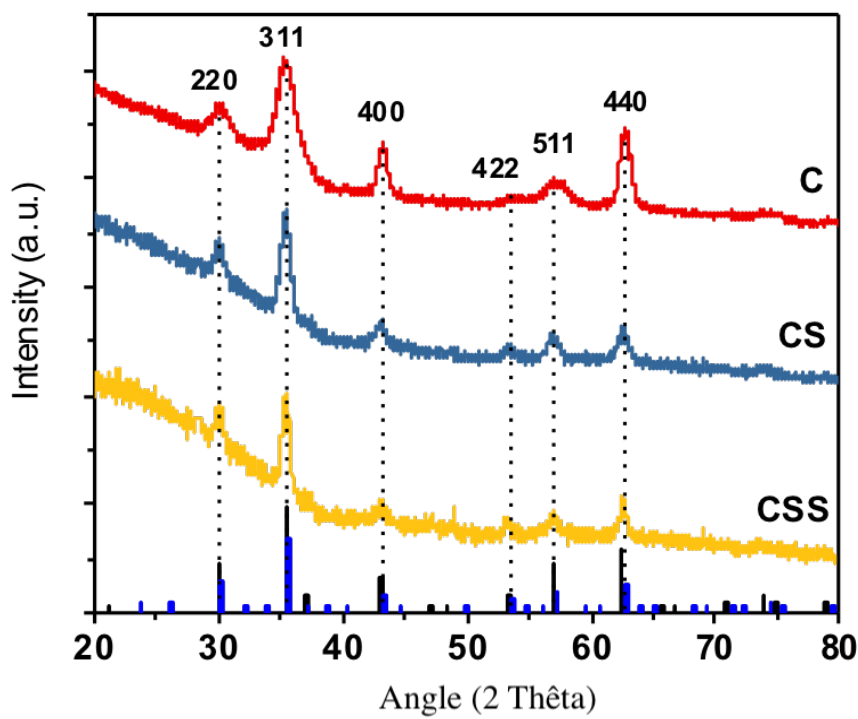

**Figure 6.** XRD patterns of C, CS and CSS nanoparticles. Black and blue bars correspond to the $Fe_3O_4$ (JCPDS card n° 19-062) and $CoFe_2O_4$ (JCPDS card n°00-022-1086) phases, respectively.

XRD patterns recorded for each nanoparticle show peaks that were attributed to the spinel structure (Fd-3m space group) (Figure 6). Unfortunately, $Fe_3O_4$ and $CoFe_2O_4$ cannot be discriminated because of similar cell parameters (a($Fe_3O_4$) = 8.396 Å, JCPDS card n°19-062 and a($CoFe_2O_4$) = 8.392 Å, JCPDS card

n°00-022-1086). Nevertheless, peaks become narrower from C, CS to CSS, which correspond to larger crystal sizes of 7.4, 10.1 and 12.0 nm (Table 1), respectively. These values are consistent with the nanoparticles' size measured from TEM micrographs. Hence confirming the good epitaxial relationship between each core and shells as observed in high resolution (HR) TEM micrographs. The slight increase of the XRD signal at low angles arises from the presence of oleic acid which is used as a ligand.

The cell parameter of C nanoparticles (8.37 Å) is intermediate to that of magnetite (a = 8.396 Å, JCPDS card n° 19-062) and maghemite (a = 8.338 Å, JCPDS card n° 39-1346) which confirms the partial oxidation of C at their surface.[51] The cell parameter of CS (8.41 Å) and CSS (8.41 Å) are larger than those of magnetite and cobalt ferrite which can be attributed to crystal strains resulting from the small size of the nanoparticles (high curvature radius). Indeed, crystal strains up to 10 % were observed by performing GPA on isolated CS and CSS nanoparticles using HR-STEM images (Figure S9). Furthermore, it may also partially account from the high content of $Fe^{2+}$.[39] These results confirm those of Lopez-Ortega *et al.*[52] who reported cell parameters of 8.40-8.42 Å for $Co_{0.6-0.7}Fe_{2.4-2.3}O_4$ nanoparticles of different sizes. They attributed such observation to the stabilization of a pure cobalt-doped magnetite phase with $Fe^{2+}$ that were not oxidized and to the presence of strains for such a small size.

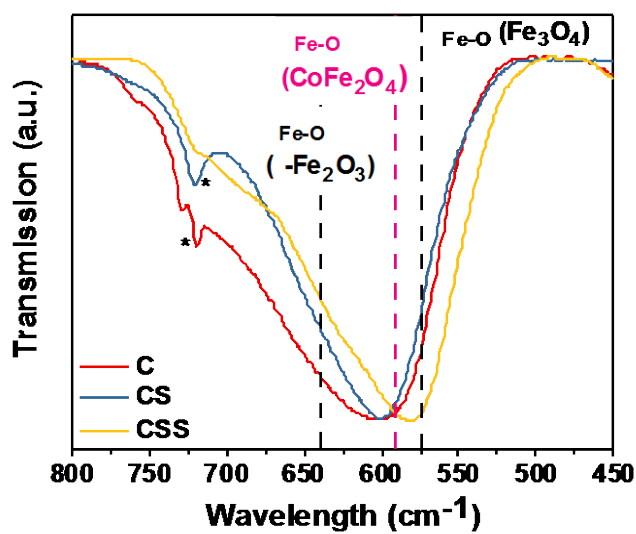

**Figure 7.** FT-IR spectra recorded at low wavelength for C, CS and CSS. Stars correspond to bands that were ascribed to some residue of iron and cobalt stearates (see Figure S10 for more information).

Fourier transform infrared (FT-IR) spectra exhibit large bands in the region from 800 to 450 cm$^{-1}$, which give additional indications on the chemical composition of the nanoparticles (Figure 7).[53] C nanoparticles display a broad band centered at 602 cm$^{-1}$ which agree with the partial oxidation of $Fe_3O_4$ in $\gamma$-$Fe_2O_3$, denoted as $Fe_{3-\delta}O_4$.[51] Indeed, $Fe_3O_4$ displays a single band at 574 cm$^{-1}$ with a shoulder at 700 cm$^{-1}$ while $\gamma$-$Fe_2O_3$ maghemite shows a maximum centered at 639 cm$^{-1}$ with several oscillations from 800 to 600 cm$^{-1}$.[53] This band shifts down to 600 cm$^{-1}$ and becomes narrower for CS, which agree with a higher content in $Fe_3O_4$. The $CoFe_2O_4$ shell partially avoids the oxidation of the core when exposed upon to air after washing. Furthermore, as cobalt ferrite displays a band at 590 cm$^{-1}$,[54] it also contributes to shift down the band. This band shifts down even lower to 581 cm$^{-1}$ for CSS, getting closer to that of magnetite (574 cm$^{-1}$). Hence, the second shell in CSS would mainly consists of magnetite, although it was expected to be fully oxidized according to our previous work on single $Fe_{3-\delta}O_4$ nanoparticles.[51] The narrower band of CS and CSS than C, and the concomitant disappearance of oscillations attributed to maghemite, confirm these observations. Thus, FT-IR shows the increase of $Fe^{2+}$ content from C, CS to CSS and the presence of $CoFe_2O_4$ in CS and CSS. Small bands around 725 cm$^{-1}$ were attributed to the H-C-H scissoring bond and correspond to negligible amounts of remaining stearates that could not be removed after washing without avoiding nanoparticle aggregation.

The chemical composition was investigated deeper by means of cationic distribution in $O_h$ and $T_d$ sites and oxidation state by performing Mössbauer and XAS/XMCD spectroscopies. $^{57}$Fe Mössbauer spectrometry brings information on the valence state of each Fe species, the local electronic structure and the magnetic environment which are described by the isomer shift δ', the quadrupolar shift ε and the hyperfine field $B_{hf}$, respectively (Table 2).

Table 2. Refined values of hyperfine parameters calculated from the fit of $^{57}$Fe Mössbauer spectra recorded at 77 K.

| Sample | isomer shift relative to α-Fe (mm/s) ±0.01 | quadrupole shift or quadrupole splitting (mm/s) ±0.01 | Hyperfine field (T) ±0.5 | Relative sub-spectral area (%) ±2 | Fe species | Site occupancy |
|---|---|---|---|---|---|---|
| C | <0.44> | <0.04> | <42.8> | 100 | $Fe^{3+}$ | - |
| CS | 0.53 | -0.04 | 53.4 | 36 | $Fe^{3+}$ | Oh |
|  | 0.41 | 0.02 | 50.9 | 51 | $Fe^{3+}$ | Td |
|  | 0.65 | 0.02 | 47.4 | 9 | $Fe^{2-3+}$ | Oh |
|  | 1.11 | 2.13 | 34.5 | 3 | $Fe^{2-3+}$ | Oh |
|  | 1.27 | 2.38 | - | 1 | $Fe^{2+}$ | |
|  | <0.50> | <0.07> | <50.5> | 99 | - | - |
| CSS | 0.50 | 0.08 | 52.5 | 46 | $Fe^{3+}$ | Oh |
|  | 0.33 | -0.07 | 51.8 | 45 | $Fe^{3+}$ | Td |
|  | 0.53 | -0.09 | 47.1 | 6 | $Fe^{3+}$ | Oh |
|  | 1.04 | 1.74 | 32.4 | 3 | $Fe^{2+}$ | Oh |
|  | <0.49> | <0.05> | <51.2> | 100 | - | - |

Mössbauer spectra were recorded at 77 K for each sample (Figure 8). They all display a resolved sextet, consistent with rather magnetic blocked state. Mössbauer spectrum of C sample displays the broadest sextet lines which are ascribed to the faster relaxation time of a fraction of spins than the measurement time of the experiment ($τ_m = 10^{-10} – 10^{-7}$ s). Such a superparamagnetic contribution can be attributed to a fraction of nanoparticle with small size (about 6.0 nm) as shown by the size distribution (Figure 2c). These results are consistent with the literature: Iron oxide nanoparticles of 11 nm measured at 77 K display no superparamagnetic contributions in Mössbauer spectra,[55] while smaller iron oxide nanoparticles of 4.6 nm measured at 77 K show significant superparamagnetic contributions.[56] The superparamagnetic contribution decreases significantly for CS (1 %) and is not observed for CSS.

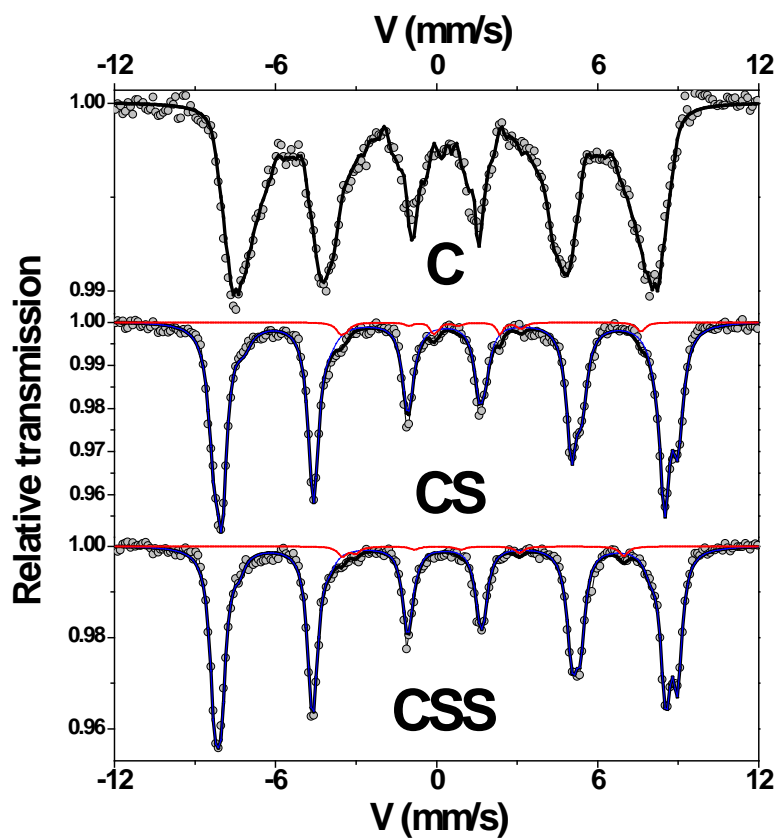

**Figure 8.** Mössbauer spectra of C, CS and CSS recorded at 77 K recorded at 77 K: black, blue and red lines correspond to the total theoretical spectrum, the total contribution of $Fe^{3+}$ and $Fe^{2-3+}$ components and the $Fe^{2+}$ components, respectively.

The fitting procedure requires great care to describe the Mössbauer spectra, which consist of magnetic sextets composed of wide, asymmetrical lines.hey have been well described by means of a discrete number of magnetic and/or quadrupolar components with independent values of isomer shift, quadrupolar shift and hyperfine field. As the solution is not unique, we report in Table 2 the corresponding refined values obtained from one example of it, but it is important to note that the mean values of hyperfine parameters are independent of the fitting procedure. Although the broad sextet lines of C were not accurate for spectrum refinement, the mean isomer shift (0.44 mm/s) is much closer to that of maghemite ($\delta$ = 0.40 mm/s) than magnetite ($\delta$ = 0.61 mm/s).[57,58] According to a linear extrapolation, C consists of approximately 19 % of magnetite and 81 % of maghemite. The mean isomer shift for CS increases to 0.50 mm/s which is correlated to a larger amount of $Fe^{2+}$ than C. More precisely, the spectrum refinement evidences two contributions with the typical $Fe^{2+}$ isomer shift (1.11 and 1.27 mm/s), the first one as a sextet corresponding to $Fe^{2+}$ in magnetically blocked magnetite, the second one as quadrupolar doublet to $Fe^{2+}$ in superparamagnetic nanoparticles containing a substantial core of magnetite. The main contribution centered at 0.53 mm/s was attributed to $Fe^{3+}$ in $O_h$ sites that account for 36 %. A second contribution centered at 0.41 mm/s was attributed to $Fe^{3+}$ in $T_d$ sites and accounts for 51 %. In addition, a third component centered at 0.65 mm/s was assigned to some intermediate $Fe^{2+,3+}$ species, which occurs below the Verwey transition. The $Fe^{3+}_{Td}/Fe^{2+,3+}_{Oh}$ ratio calculated for CS is approximately 1, which is much higher than the theoretical value of 0.5 for pure magnetite. Considering the core-shell structure where $Co^{2+}$ partially replaces $Fe^{2+}$ in Oh sites, an intermediate value was expected. It may be attributed to a super stoichiometry in oxygen or to the presence of vacancies.[59]

Furthermore, high hyperfine fields of 53.4 T and 50.9 T for $Fe^{3+}$ in $O_h$ and $T_d$ sites, respectively, agree with the presence of Co species in the vicinity of Fe species, consistent with $CoFe_2O_4$.[60,61] In contrast, hyperfine fields of 47.4 and 34.5 T measured for $Fe^{3+}$ and $Fe^{2+}$ in $O_h$ sites correspond to the $Fe_{3-\square}O_4$ core. According to isomer shifts reported previously for $Fe_{2.95}O_4$ (0.61 mm/s)[62] and $CoFe_2O_4$ (0.45 mm/s)[63] nanoparticles, the mean isomer shift of CS (0.51 mm/s), would correspond to a composition of 63 % of $CoFe_2O_4$ and 37 % of $Fe_{3-d}O_4$, i.e. a core size of 6.8 nm and a shell thickness of 1.6 nm, assuming a simple core@shell model with a radial structure. According to the size variation measured from TEM micrographs, we expected a cobalt ferrite shell thickness of 1 nm. Therefore, such a thicker cobalt ferrite shell may result from the partial solubilisation of the iron oxide core followed by the recrystallization[35] of Fe monomers with Co monomers in a similar way we reported earlier.[23,39,40] Considering that a stoichiometric ratio of Fe and Co stearate was used, we expect the cobalt ferrite shell to be under stoichiometric.

In CSS, the mean isomer shift decreased slightly to 0.49 mm/s which evidences a slightly lower content of $Fe^{2+}$ than in CS. The contribution centered at 1.04 mm/s felt down to 3 %. Additional contributions centered at 0.53 mm/s and 0.50 mm/s were attributed to $Fe^{3+}$ in $O_h$ sites (6 % and 46 %, respectively). A third contribution centered at 0.33 mm/s was attributed to $Fe^{3+}$ in $T_d$ sites (45 %). The lower $Fe^{3+}_{Td}/Fe^{2+,3+}_{Oh}$ ratio (0.82) for CSS than for CS (1) agrees with higher amount of $Fe^{2+}$ that may be localized in the second shell of iron oxide that was grown at the surface of the $CoFe_2O_4$ shell. High hyperfine fields (52.5 and 51.8 T) calculated for $Fe^{3+}$ in $O_h$ and $T_d$ sites also confirm the presence of cobalt ferrite within CSS. The slight increase of their sub spectral areas in comparison with CS is indicative of a larger fraction of cobalt ferrite. It is clearly confirmed by the higher mean isomer shift which corresponds to 75 % of cobalt ferrite and 25 % of $Fe_{3-\square}O_4$. Thus, the composition of CSS nanoparticles would consist in a core of 6.8 nm, a 2.9 nm thick cobalt ferrite shell and a 0.3 nm thick $Fe_{3-\square}O_4$ shell, which is in agreement with STEM-EELS measurements. Therefore, the cobalt ferrite shell in CSS would be thicker than in CS (1.6 nm).

Mössbauer spectrometry has shown that the $Fe^{2+}$ content increases from C to CS but slightly decreases from CS to CSS. It has also shown that the $Fe_{3-\square}O_4$ core size decreases concomitantly to the cobalt ferrite shell which increases further from CS to CSS thus resulting in a much thinner $Fe_{3-\square}O_4$ shell than expected. According to the results obtained with the above-mentioned techniques (the increase of the mean cell parameter (XRD) and the shift of the M-O band to lower frequencies (FT-IR) from C, CS to CSS as well as the broad Co distribution in CSS nanoparticles revealed by STEM-EELS (mapping and spectra), it seems that the evolution of the cobalt ferrite phase predominates over the variation of the $Fe^{2+}$ content between CS and CSS.

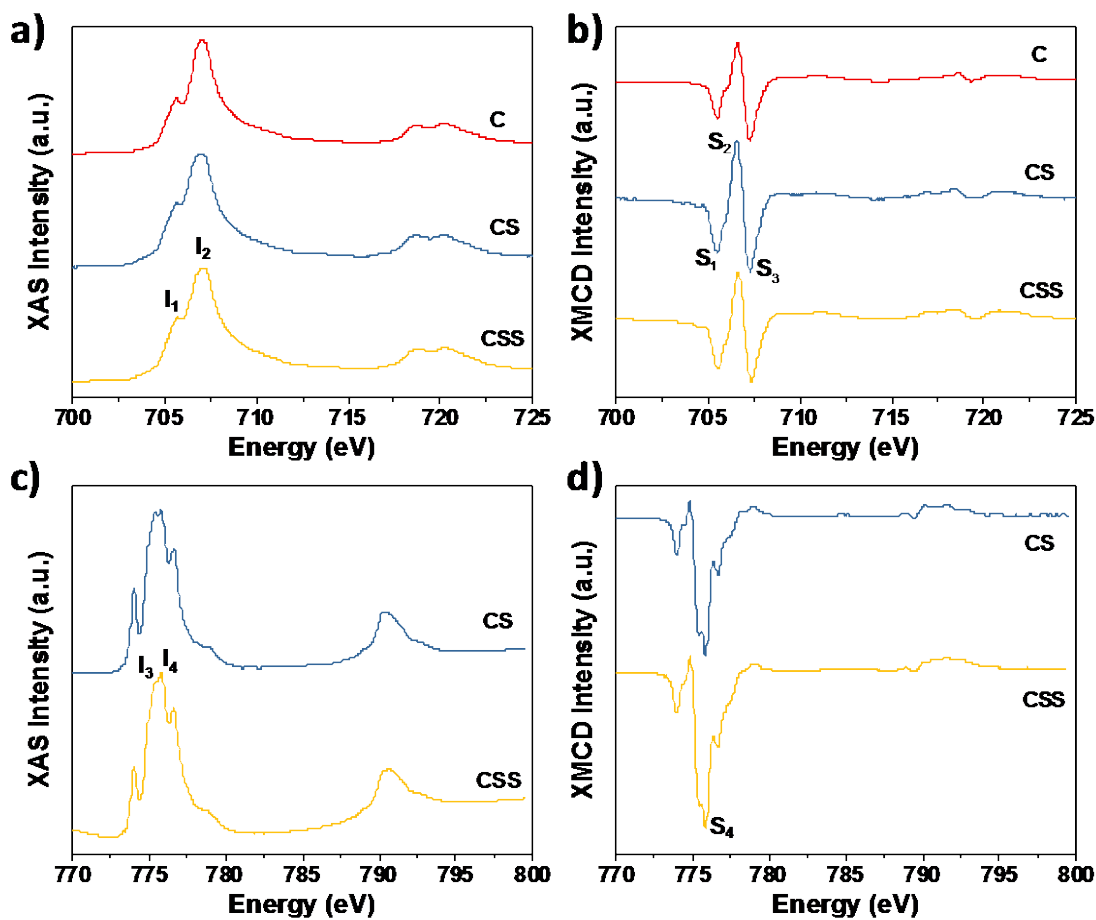

**Figure 9.** a), c) Isotropic XAS and b, d) XMCD spectra at the a), b) Fe $L_{2,3}$ edges and at the c), d) Co $L_{2,3}$ edges of C, CS and CSS nanoparticles.

The isotropic XAS and XMCD spectra recorded at the Fe $L_{2,3}$ edges (Figure 9a and b) are all typical of a spinel ferrite structure.[64–66] XAS spectra evidenced two main contributions in the $L_3$ region that were ascribed to $Fe^{2+}$ in $O_h$ sites (peak $I_1$) and to $Fe^{3+}$ in $O_h$ and $T_d$ sites (peak $I_2$). Hence, the intensity ratio $I_1/I_2$ reported as 0.71[67] for $Fe_3O_4$ and 0.35 for $\gamma$-$Fe_2O_3$[67] brings further information on the $Fe^{2+}$ content within the nanoparticles. The value calculated for C (0.53) agrees with an intermediate composition of between magnetite and maghemite. Then, it increases for CS (0.64) corresponding to a higher content of $Fe^{2+}$. These results agree with those of XRD, FT-IR spectroscopy and Mössbauer spectrometry and those of our previous work on similar core-shell nanoparticles.[23] The $I_1/I_2$ ratio slightly decreases for CSS (0.62) which is ascribed to a slightly lower content in $Fe^{2+}$ (as observed from Mössbauer spectrometry).

XMCD spectra recorded at the Fe $L_{2,3}$ edges display three main peaks in the $L_3$ region where the S1 and S3 peaks were respectively attributed to $Fe^{2+}$ and $Fe^{3+}$ in $O_h$ sites, while the S2 peak corresponds to $Fe^{3+}$ in $T_d$ sites, which spins are coupled antiparallel to Fe cations in $O_h$ sites. Such consideration is typical of the ferrimagnetic coupling of Fe spins in the reverse spinel structure of iron oxide and cobalt ferrite. The intensity ratio S=(S1+S2)/(S2+S3) brings further information on the oxidation state of iron cations. Hence, magnetite displays a higher ratio (1.14) than maghemite (0.69).[67] C sample displays the closest ratio (0.77) to maghemite, while it increases for CS (0.85) and get even higher for CSS (0.90). Such an increase of $Fe^{2+}$ content from CS to CSS is contradictory with the $I_1/I_2$ ratio measured from XAS spectra and to Mössbauer spectrometry. XMCD being a polarized mode, it may favor $Fe^{2+}$ uncompensated spins at the nanoparticle surface resulting from the break of symmetry vs. $Fe^{3+}$ spins which are coupled antiparallel. In addition, we observed an excess of $Fe^{3+}$ in $T_d$ sites in the XMCD spectra of CS and CSS (Figure S13) which may result from a preferential occupancy of the $O_h$ sites by the $Co^{2+}$ cations in the cobalt ferrite structure.

Isotropic XAS spectra recorded at the Co $L_{2,3}$ edges confirmed the presence of $Co^{2+}$ in $O_h$ sites of a spinel structure.[66,68] The $I_4$ peak is slightly higher than the $I_3$ peak for CS which is more obvious for CSS. Such observation qualitatively shows the increase of the cobalt ferrite content[23,69] from CS to CSS nanoparticles. The XMCD spectra recorded at the Co $L_{2,3}$ edges are all typical of $Co^{2+}$ cations in $O_h$ sites.[68,70,23] All spectra being normalized at the edge of the energy jump, the intensity of the S4 peak (95 % for CS and 108 % for CSS with respect to XAS signal) agree with an increase of uncompensated $Co^{2+}$ spins from a $CoFe_2O_4$ phase between CS and CSS samples.[71,44] It confirms the increase of $Fe^{2+}$ from C, CS to CSS and XAS and XMCD spectra unambiguously show the presence of a $CoFe_2O_4$ phase, which increases from CS to CSS.

Element-specific magnetization curves were recorded at the Fe S2, S3 and Co S4 peak energies for CS, and at the Fe S2 and Co S4 peak energies for CSS (Figure 10). The selective hysteresis curves recorded for CS at different energies showed similar coercive fields ($H_C$) of about 6.5 kOe (Table 3). It shows that Fe spins in $T_d$ and $O_h$ sites and Co spins in $O_h$ sites are magnetically coupled which confirms the presence of $CoFe_2O_4$.[23,44] CSS behaves similarly although $H_C$ is larger (about 9.5 kOe), agreeing with a thicker cobalt ferrite layer than for CS, as shown by Mössbauer spectrometry. These values may differ from those obtained by magnetometry (see below) because the sample's preparation is different, which induces some variations of dipolar interactions between nanoparticles.

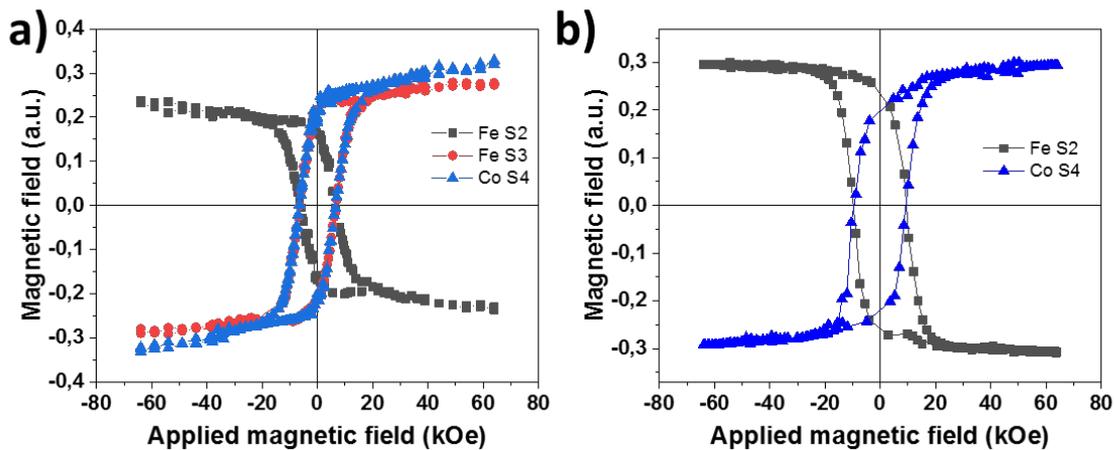

**Figure 10.** Element-specific magnetization curves recorded at 4 K by XMCD at the Fe and Co $L_{2,3}$ edges in a) CS, b) CSS.

**Table 3.** Magnetic characteristics of element specific magnetization curves.

| Sample | $H_C$ Fe S2 (kOe) | $H_C$ Fe S3 (kOe) | $H_C$ Co S4 (kOe) | $<H_C>$ (kOe) |
|---|---|---|---|---|
| **CS** | 6,3 | 6,7 | 6,6 | **6,5 ±** |
| **CSS** | 9,7 | - | 9,4 | **9,5 ±** |

The magnetic properties of C, CS and CSS were investigated by SQUID magnetometry (Figure 11). Magnetization curves recorded against temperature (M(T)) after zero field cooling (ZFC) show a maximum at $T_{max}$ which is usually assimilated to the transition temperature between blocked magnetic moments and the superparamagnetic behavior. $T_{max}$ measured for C (86 K) agrees with values reported for iron oxide nanoparticles of similar sizes.[51] It increases for CS (290 K), and further for CSS (300 - 350

K), agreeing with higher magnetic anisotropy energy ($E_a$) which results from the modification of the nanoparticle structure. Although M(T) curves recorded for C are very typical of magnetic iron oxide nanoparticles, the ones recorded for CS and CSS correspond to different magnetic properties. For CS, a kink at temperatures slightly below $T_{max}$ refers to a minor fraction of nanoparticles which exhibit a lower effective magnetic anisotropy energy than the rest of the sample. As long as the size distribution of CS is rather narrow, it may results from inhomogeneous spatial distribution and weaker dipolar interactions which may be partially due to the oleic acid.[72] For CSS, the increase of magnetization at temperature higher than $T_{max}$ can be ascribed to super magnetic domains resulting from strong dipolar interactions between nanoparticles, e.g. superferromagnetism.[2,73] Indeed, the M(T) field cooled (FC) curve show almost constant magnetization at low temperatures which agrees with dipolar interactions between nanoparticles.[74,75] Therefore, $T_{max}$ corresponds to a broad distribution of temperatures which is difficult to assess precisely.

The transition between blocked and flipped magnetic moments is more accurately described by the blocking temperature ($T_B$) which corresponds to a distribution of energy barriers. $T_B$ can be easily extracted from the ZFC-FC M(T) curves using the following equation:[76]

$$f(T_B) = - \frac{[dM_{ZFC} - M_{FC}]}{T[dT]} \qquad (1)$$

$T_B$ corresponds to temperature distributions centered at 48, 239 and 280 K for C, CS and CSS, respectively (Figure 11b). The dramatic enhancement of $T_B$ from C to CS is attributed to arise from interfacial exchange coupling between the soft $Fe_{3-\square}O_4$ and the hard $CoFe_2O_4$ phases. Temperature distributions of CS and CSS are more complex than that of C. For CS, an additional contribution centered at 278 K corresponds to the kink observed in M(T) curves. CSS displays a minor contribution centered at 390 K which may be attributed to the presence of super ferromagnetic domains.[2,73] A second contribution centered at 220 K can be attributed to CSS nanoparticles which are partially or not covered by a second shell.

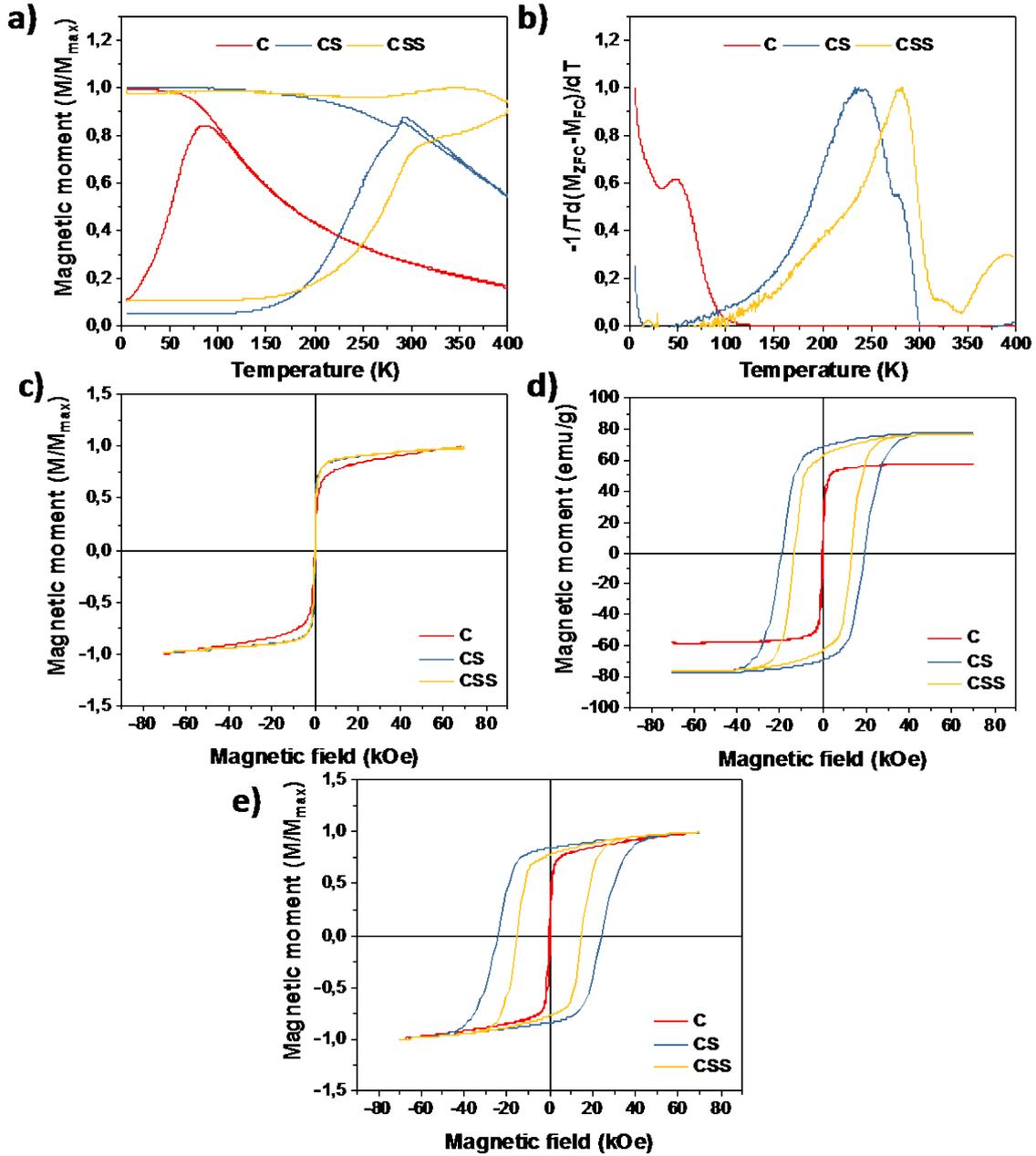

**Figure 11.** Magnetic characterizations of C, CS and CSS. a) Magnetization curves recorded against temperature after zero field cooling (ZFC) and field cooling (FC). b) Distribution of blocking temperatures ($T_B$). Magnetization measurements recorded against a magnetic field at c) 300 K, d) 5 K after ZFC, and e) 10 K after FC under 7 T.

Considering the anisotropy constants ($K(CoFe_2O_4)^{8,77} \approx 1\text{–}6.10^5$ J/m$^3$ and $K(Fe_3O_4)^{73} \approx 1\text{–}5.10^4$ J/m$^3$) and the volume V of each phase,

$$K(CoFe_2O_4).V(CoFe_2O_4) \gg K(Fe_3O_4).V(Fe_3O_4) \qquad (2)$$

Therefore, in exchange coupled nanoparticles, the effective magnetic anisotropy ($E_{eff}$) of CS and CSS can be assimilated to:[78]

$$E_{eff} = K_{eff}V = K(CoFe_2O_4).V(CoFe_2O_4) = 25K_BT_B \qquad (3)$$

with V the total volume of the nanoparticle, and $k_B$ the Boltzmann constant. $K_{eff}$ calculated for CS (1.58 $10^5$ J/m$^3$) agrees with the anisotropy constants reported for CoFe$_2$O$_4$ nanoparticles. Then, $K_{eff}$ significantly decreases for CSS (8.2 $10^4$ J/m$^3$) and gets very close to C (6.2 $10^4$ J/m$^3$) although the soft-shell volume is low (150 nm$^3$) and the volume of the CoFe$_2$O$_4$ shell is increased by four times. According to the anisotropy constant calculated for CS, a much higher $T_B$ (400 K) was expected for CSS. In contrast, the increase of $T_B$ (40 K) from CS to CSS is higher than that (20 K) corresponding to the volume of the Fe$_{3-\delta}$O$_4$ shell. Therefore, the increase of $T_B$ is not ascribed to the volume increase but to exchange coupling at the CoFe$_2$O$_4$/Fe$_{3-\delta}$O$_4$ that contributes to the enhancement of the effective magnetic anisotropy energy of nanoparticles.

$H_C$ temperature dependent curves of CS and CSS (Figure S12) also bring information on the magnetic behavior of CS and CSS. $H_C$ decreases with increasing the temperature until the onset at similar temperatures (about 265 K) for both CS and CSS (Figure S12). Nevertheless, $H_C$ is higher for CS than for CSS and decreases faster when the temperature rises up. According to the Stoner-Wohlfarth model,[79] it is ascribed to a higher effective magnetic anisotropy energy for CS than for CSS.

$$H_C = 0.48 H_K \left[1 - \left(\frac{T}{T_B}\right)^{0.5}\right] \quad (4)$$

with the anisotropic field $H_K = \frac{2K_{eff}}{M_S}$

Magnetization curves recorded against an applied magnetic field at 300 K perfectly overlap for each sample which agree with superparamagnetic behavior. In contrast, M(H) curves recorded at 5 K show opened hysteresis corresponding to blocked magnetic moments. The coercive field ($H_C$) measured for C (300 Oe) is typical of iron oxide nanoparticles.[51] It increases dramatically for CS (19.2 kOe) because of the strong interfacial exchange-coupling between the hard CoFe$_2$O$_4$ shell and the soft Fe$_{3-\delta}$O$_4$ core.[1] Then, it decreases for CSS (13.1 kOe) because of the presence of the soft Fe$_{3-\delta}$O$_4$ shell. It is consistent with the decrease of the effective magnetic anisotropy constant in comparison with CS, as observed for CoFe$_2$O$_4$@MnFe$_2$O$_4$ core-shell nanoparticles.[24] Same trends were observed for the remanent magnetization ($M_R$) and the $M_R/M_S$ ratio (Table 4). The volume of the Fe$_{3-\delta}$O$_4$ shell (160 nm$^3$) is much smaller than that corresponding to the increase of the CoFe$_2$O$_4$ shell (500 nm$^3$). Therefore, $H_C$ being dependent on the fractions of hard and soft phases,[80] it was expected to increase. Indeed, the Fe$_{3-\delta}$O$_4$ phase favors the magnetic reversal of interfacial spins of CoFe$_2$O$_4$ which results in lower $H_C$. Furthermore, no kinks were observed in M(H) curve, agreeing with an effective exchange coupling between both Fe$_{3-\delta}$O$_4$ and CoFe$_2$O$_4$ phases, that propagate through the entire volume of nanoparticles, whatever the interface in both CS or CSS structures.

M(H) curves recorded for CS and CSS after field cooling (7 T) from 300 K to 10 K show larger $H_C$ than ZFC M(H) curves. It is ascribed to the alignment of soft spins with the applied magnetic field that favors their coupling with hard spin, thus resulting in magnetic reversal at higher magnetic fields. Moreover, the hysteresis curves were not shifted to low magnetic fields. Such a behavior being typical of exchange bias coupling between soft FiM and hard antiferromagnetic phases, it agrees with the absence of CoO.[14,39,81]

The increase of saturation magnetization ($M_S$) from C (58 emu/g) to CS (78 emu/g) and CSS (77 emu/g) is also indicative of their chemical composition which agree with a thicker CoFe$_2$O$_4$ shell.[50,60] Indeed, for similar sizes, CoFe$_2$O$_4$ nanoparticles[84,85] display a higher $M_S$ than Fe$_{3-\delta}$O$_4$ nanoparticles.[42] Although these values are lower than bulk values (Fe$_3$O$_4$ : 98 emu/g, CoFe$_2$O$_4$ : 94 emu/g) because of surface effects,[42] the higher $M_S$ of CS and CSS than C agrees with the coupling of the Fe$_{3-\delta}$O$_4$ interfacial spins by that of the CoFe$_2$O$_4$. The slight decrease of $M_S$ from CS to CSS agrees with the formation of a very thin shell of Fe$_{3-\delta}$O$_4$.

Table 4. Magnetic characteristics of C, CS and CSS.

| | C | CS | CSS |
|---|---|---|---|
| Size (nm) | 8.0 | 10.0 | 13.1 |
| Shell thickness (nm) | - | 1.0 | 1.5 |
| $H_C$ 5 K (ZFC) kOe | 0.3 | 19.2 | 13.1 |
| $H_C$ 10 K (FC) kOe | 0.3 | 24.1 | 15 |
| $H_E$ 10 K (FC) Oe | 0 | 0 | 0 |
| $T_{max}$ (K) | 86 | 290 | 301-400 |
| $T_B$ (K) | 48 | 239 | 280 / 327 |
| $K_{eff}$ ($10^4$ J.m$^{-3}$) | 6.2 | 15.8 | 8.2 |
| $M_S$ 5K (ZFC) emu/g | 58 | 78 | 77 |
| $M_R$ 5K (ZFC) emu/g | 15 | 69 | 63 |
| $M_R/M_S$ | 0.26 | 0.89 | 0.82 |

# Discussion

TEM micrographs showed that the size of the nanoparticles increased from 8.0 ± 0.9 nm (C) to 10.0 ± 1.5 nm (CS) and, further to 13.1 ± 2.2 nm (CSS), in agreement with the successive growth of shells. Nevertheless, the complexity of crystal growth processes resulted in broader size distributions and in shape deviation from spheres. It can be ascribed to various parameters such as kinetics (reagent concentration, temperature, mass transport, capping agent) and thermodynamics (energy barrier, surface energy) parameters.[86] Iron oxide nanoparticles exhibit a faceted shape that consists in the {100} and {110} planes as usually observed for cubic crystallographic structures.[31,42] *hkl* reflections featuring different surface energies, they may favor heterogeneous seed-mediated growth. Other processes such as the selective binding of oleic acid acting as capping agent on specific {*hkl*} planes and the competition of adsorption vs. diffusion of atoms on crystal surface markedly alter crystal growth.[86]
HAADF-HRSTEM micrographs revealed a single crystal-like structure for each nanoparticle, thanks to the negligible lattice mismatch between both spinel structures. FFT calculated from HR-TEM micrographs recorded on the edge and at the center of the nanoparticles evidenced the good epitaxial relationship of the shells with the core. These results were confirmed by the increase of the crystal size from C to CS and further to CSS, as it has been measured from XRD patterns. The cell parameter calculated for the iron oxide core corresponds to partially oxidized $Fe_{3-\square}O_4$ nanoparticles which agrees with the literature. Values calculated for CS and CSS are higher than that of C and correspond to higher $Fe^{2+}$ content and to the growth of a cobalt ferrite shell. However, they are larger than the theoretical value of cobalt ferrite which is related to the presence of crystal strains, as shown by GPA.[84] Slight stacking defaults in the crystal structure were also observed from HR-TEM micrographs at the edge of the nanoparticles.

Depending on the chemical composition of nanoparticles, since XRD cannot accurately discriminate $Fe_{3-\square}O_4$ from $CoFe_2O_4$ due to the very similar cell parameters of the two crystalline structures, a wide set of complementary analysis techniques was used. The EELS-SI mapping performed on individual nanoparticles clearly showed the presence of Co atoms at the edges of the CS nanoparticles. In CSS, EELS data suggest that the Co atoms are not confined to the first shell, but are distributed in the core and in the second shell. Although EELS cross-sections performed at different position – i.e. at the center and on the edges - of CSS did not show a decrease in the Co content at the very edge of CSS, EDX measurements performed on groups of nanoparticles showed that the Fe:Co atom ratio increased from CS to CSS which is consistent with the growth of an iron oxide shell at the surface of CS. The FT-IR spectra gave more details on the evolution of the chemical composition of the nanoparticles. The spectra showed that the M-O vibration band became narrower and was shifted to lower frequencies from C to CS and, further to CSS, which is consistent with the formation of cobalt ferrite as well as the increase of $Fe^{2+}$ content after the growth of each shell.

The chemical structure of the nanoparticles was investigated more deeply by performing Mössbauer spectrometry and XAS/XMCD spectroscopy. Mössbauer spectrometry gave information on the oxidation state and the site occupancy of Fe atoms. The spectra showed that the content of $Fe^{2+}$ increased from C nanoparticles to CS which is consistent with the preservation of the core against oxidation upon exposure to air through the formation of the cobalt ferrite shell. The increase of the average hyperfine field with respect to that of $Fe_{3-\square}O_4$ nanoparticles is indicative of a significant fraction of $Fe^{3+}$ cations in a cobalt ferrite structure. Considering the respective fractions of the different cations (oxidation state and site occupancy), CS consists of a core of 6.8 nm, surrounded by a 1.6 nm thick $CoFe_2O_4$ shell. Such a reduction of the iron oxide core (8.0 nm) while the size of the nanoparticles increases from C to CS, was ascribed to the partial solubilisation-recrystallization process upon heating at high temperature.[35]

Fe monomers generated by the partial solubilisation of the $Fe_{3-\square}O_4$ core contribute to the formation of the $CoFe_2O_4$ shell. Although the $CoFe_2O_4$ shell becomes thicker, Fe and Co stearates being added as stoichiometric, the $CoFe_2O_4$ shell is certainly sub stoichiometric and corresponds to about $Co_{0.79}Fe_{2.21}O_4$. In CSS, the $CoFe_2O_4$ shell became thicker (up to 2.9 nm) than CS because some Co stearate that remains in the nanoparticle suspension (see FTIR, Figure 8) reacted with Fe stearate that was added in the reaction medium. Therefore, most of Fe monomers contribute to the extension of the cobalt ferrite shell, while a small fraction led to the formation of the second $Fe_{3-\square}O_4$ shell which is very thin (0.3 nm). This value being smaller than the cell parameter, such a shell did not grew homogeneously at the surface of CSS. These results were confirmed by STEM-EELS line profiles performed across CS and CSS nanoparticles (Figures S2 and S3). These STEM-EELS analyses showed that the Co content exists in larger volume than what was expected from the size variation measured from TEM micrographs. Furthermore, the Co content increases at the edge of CS as expected (Figure S4). It is much lower for CSS, which agree with the formation of a very thin shell of $Fe_{3-\square}O_4$ although it was expected to disappear (Figure S4).

The XAS and XMCD spectra recorded at the Fe and Co edges provide additional information on the chemical structure of the nanoparticles. The evolution of several peaks in XAS spectra at Fe $L_3$ edge agrees with an intermediate phase between magnetite and maghemite. It confirmed that the $Fe^{2+}$ content increased from C to CS, and slightly decreased in CSS, in accordance with XRD, FT-IR spectroscopy and Mössbauer spectrometry. XMCD spectra are signatures of a ferrimagnetic coupling of Fe spins in the reverse spinel structure of iron oxide and cobalt ferrite. However, XMCD shown a weak increase of $Fe^{2+}$ content from CS to CSS which may be due $Fe^{2+}$ uncompensated spins at the nanoparticle surface resulting from the break of symmetry vs. $Fe^{3+}$ spins in $O_h$ and $T_d$ sites which are coupled antiparallel. Furthermore, XAS and XMCD spectra at the Co $L_3$ edge unambiguously demonstrate the presence of $CoFe_2O_4$ in CS and CSS with a $Co^{2+}$ cations in $O_h$ sites. The magnitude of XMCD confirms the increase of the cobalt ferrite phase form CS to CSS. This result was confirmed by element-specific magnetization curves recorded at Fe and Co edges. Similar $H_C$ agree with strong exchange interactions between Fe and Co spins in a spinel structure.

These results show that the chemical composition and the structure of CS and CSS (Figure 12) differ significantly from what was expected from the size variation of the nanoparticles measured from TEM micrographs and the Fe/Co molar ratio calculated from EDX analysis.

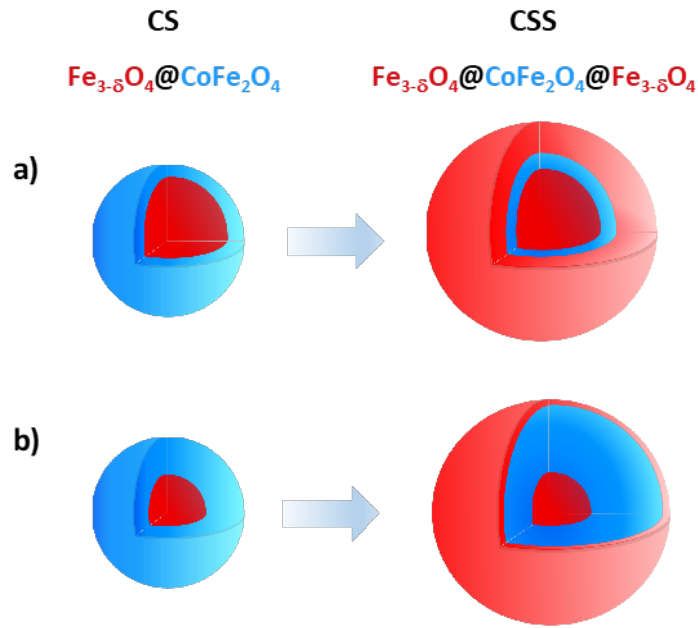

**Figure 12.** Schematic illustration of the nanoparticle structure a) expected from size variation measured from TEM micrographs and Fe/Co molar ratio measured from EDX, and b) from data deduced from Mössbauer and XAS/XMCD spectra.

The magnetic properties of CS and CSS are directly correlated to their chemical composition and crystal structure. The significant increase of $T_B$ from C to CS, and further to CSS agrees with strong hard-soft magnetic exchange coupling at $Fe_{3-\delta}O_4$/$CoFe_2O_4$ and $CoFe_2O_4$/$Fe_{3-\delta}O_4$ interfaces. The highest $T_B$ value of 280 K is very close to that (300 K) measured for $Co_{0.68}Fe_{2.32}O_4$ nanoparticles with similar size.[85] It shows that interfacial exchange coupling mostly compensates the lack of magnetic anisotropy energy corresponding to the volume of the $Fe_{3-\delta}O_4$ core. The contribution of the exchange coupling phenomenon at the soft/hard interface to $T_B$ is more significant than that of the hard/soft shell. It can be ascribed to the inhomogeneous coating of the $Fe_{3-\delta}O_4$ shell which limits the hard/soft interface and the coupling efficiency. Nevertheless, the combination of soft/hard and hard/soft interfaces contributes to the enhancement of the effective magnetic anisotropy energy. Furthermore, the smooth variation of magnetization in each M(H) curves (no kinks) confirms strong hard-soft exchange coupling at core/shell and shell/shell interfaces which propagates efficiently through the entire core-shell structure. The decrease of $H_C$ from CS to CSS is also ascribed to the growth of the second $Fe_{3-\delta}O_4$ shell. The increase of the hard $CoFe_2O_4$ shell volume from CS to CSS has no significant effect on these parameters. In contrast, $M_S$ increased significantly from C to CS. Interestingly, CSS displays much stronger dipolar interactions than CS although they display similar $M_S$. It means that the super ferromagnetic behavior is significantly dependent on the amount of oleic acid grafted at the nanoparticle surface which control interparticle distances.

Conclusion

$Fe_{3-\delta}O_4$@$CoFe_2O_4$@$Fe_{3-\delta}O_4$ (CSS) nanoparticles were synthesized by the thermal decomposition method through two successive seed-mediated growth steps. The chemical composition and the crystal structure of the nanoparticles were investigated by a wide set of analysis techniques. First of all, the thickness of the shells in CS and CSS (Figure 12) vary significantly from values expected from TEM micrographs and EDX analysis. Indeed, the reaction mechanism consists in the partial solubilisation of the nanoparticles followed by the recrystallisation of monomers which results in significantly different spatial distributions of Fe and Co cations in the nanoparticle volume. While such a mechanism leads to a thicker cobalt ferrite shell, it is under stoichiometric with respect to Co content. In contrast, the second iron oxide shell is so thin that it does not homogeneously cover the cobalt ferrite shell. Nevertheless, each nanoparticle exhibits a single crystal-like structure because of the very low lattice mismatch between both spinel phases. Some crystal defects which arise from the nanoparticle geometry and the different surface energy of facets could be observed. Although cobalt ferrite could not be discriminated from iron oxide by XRD, its presence was unambiguously showed by XAS/XMCD measurements and confirmed by Mössbauer spectrometry and FT-IR spectroscopy. The $CoFe_2O_4$ shell avoids the oxidation of $Fe^{2+}$ at the surface of the iron oxide core which results in a higher $Fe^{2+}$ content in CS that slightly decreases in CSS.

CS and CSS nanoparticles display enhanced magnetic anisotropy energies in comparison to C as a result of strong exchange coupling at the soft/hard and hard/soft interfaces. Although the $Fe_{3-\delta}O_4$ shell is very thin and discontinuous, it has a significant influence on the magnetic properties of CS nanoparticles. In comparison, pure $CoFe_2O_4$ nanoparticles with similar size to that of CSS display close value of $T_B$ to ours, although a much lower amount of Co atoms was incorporated. Furthermore, the softness of the $Fe_{3-\delta}O_4$ shell resulted in the decrease of $H_C$ and $M_R$ while $M_S$ was preserved. Such a high control of the structure of the nanoparticles is particularly interesting to modulate their magnetic properties, thus extending their potential to a wide range of applications.

## ASSOCIATED CONTENT

**Supporting Information.** Additional data related to STEM-HAADF micrographs, EELS spectra, EELS-SI mapping, GPA, FTIR, granulometry, magnetism, XMCD spectra. The Supporting Information is available free of charge on the ACS Publications website.

## AUTHOR INFORMATION

**Corresponding Author**

benoit.pichon@ipcms.unistra.fr

**Notes**

The authors declare no competing financial interests.

## ACKNOWLEDGMENT

K.S. was supported by a PhD grant from the French Agence Nationale de la Recherche (ANR) under the reference ANR-11-LABX-0058-NIE within the Investissement d'Avenir program ANR-10-IDEX-0002-02 and SOLEIL synchrotron / Laboratoire Léon Brillouin fellowship. The authors are grateful to SOLEIL synchrotron light source for providing the access to DEIMOS beamline and to DIAMOND synchrotron light source for providing access to BLADE beamline. HR-STEM and STEM-EELS studies were conducted at the Laboratorio de Microscopias Avanzadas, Universidad de Zaragoza, Spain. S.H. is grateful to DFG (HE 7675/1-1). R.A. gratefully acknowledges the support from the Spanish Ministry of Economy and Competitiveness (MINECO) and the MICINN through project grants MAT2016-79776-P (AEI/FEDER, UE) and PID2019-104739GB-100/AEI/10.13039/501100011033 as well as from the European Union H2020 program "ESTEEM3" (823717).

# References

(1)     López-Ortega, A.; Estrader, M.; Salazar-Alvarez, G.; Roca, A. G.; Nogués, J. Applications of Exchange Coupled Bi-Magnetic Hard/Soft and Soft/Hard Magnetic Core/Shell Nanoparticles. *Physics Reports* **2015**, *553*, 1–32.

(2)     Bedanta, S.; Kleemann, W. Supermagnetism. *J. Phys. D: Appl. Phys.* **2008**, *42* (1), 013001.

(3)     Massari, S.; Ruberti, M. Rare Earth Elements as Critical Raw Materials: Focus on International

Markets and Future Strategies. *Resources Policy* **2013**, *38* (1), 36–43.
(4)     Goll, D.; Loeffler, R.; Herbst, J.; Karimi, R.; Schneider, G. High-Throughput Search for New Permanent Magnet Materials. *J. Phys.: Condens. Matter* **2014**, *26* (6), 064208.
(5)     Kuz'min, M. D.; Skokov, K. P.; Jian, H.; Radulov, I.; Gutfleisch, O. Towards High-Performance Permanent Magnets without Rare Earths. *J. Phys.: Condens. Matter* **2014**, *26* (6), 064205.
(6)     Tartaj, P.; Morales, M. P.; Gonzalez-Carreño, T.; Veintemillas-Verdaguer, S.; Serna, C. J. The Iron Oxides Strike Back: From Biomedical Applications to Energy Storage Devices and Photoelectrochemical Water Splitting. *Advanced Materials* **2011**, *23* (44), 5243–5249.
(7)     Cheon, J.; Park, J.-I.; Choi, J.; Jun, Y.; Kim, S.; Kim, M. G.; Kim, Y.-M.; Kim, Y. J. Magnetic Superlattices and Their Nanoscale Phase Transition Effects. *PNAS* **2006**, *103* (9), 3023–3027.
(8)     Lopez-Ortega, A.; Estrader, M.; Salazar-Alvarez, G.; Estrade, S.; Golosovsky, I. V.; Dumas, R. K.; Keavney, D. J.; Vasilakaki, M.; Trohidou, K. N.; Sort, J.; Peiró, F.; Suriñach, S.; Baró, M. D.; Nogués J. Strongly Exchange Coupled Inverse Ferrimagnetic Soft/Hard, MnxFe3-XO4/FexMn3-XO4, Core/Shell Heterostructured Nanoparticles. *Nanoscale* **2012**, *4* (16), 5138.
(9)     Lima, E.; Winkler, E. L.; Tobia, D.; Troiani, H. E.; Zysler, R. D.; Agostinelli, E.; Fiorani, D. Bimagnetic CoO Core/CoFe$_2$O$_4$ Shell Nanoparticles: Synthesis and Magnetic Properties. *Chemistry of Materials* **2012**, *24* (3), 512–516.
(10)    Lavorato, G. C.; Lima, E.; Troiani, H. E.; Zysler, R. D.; Winkler, E. L. Tuning the Coercivity and Exchange Bias by Controlling the Interface Coupling in Bimagnetic Core/Shell Nanoparticles. *Nanoscale* **2017**, *9* (29), 10240–10247.
(11)    Lottini, E.; López-Ortega, A.; Bertoni, G.; Turner, S.; Meledina, M.; Van Tendeloo, G.; de Julián Fernández, C.; Sangregorio, C. Strongly Exchange Coupled Core|Shell Nanoparticles with High Magnetic Anisotropy: A Strategy toward Rare-Earth-Free Permanent Magnets. *Chem. Mater.* **2016**, *28* (12), 4214–4222.
(12)    Manna, P. K.; Yusuf, S. M.; Basu, M.; Pal, T. The Magnetic Proximity Effect in a Ferrimagnetic Fe$_3$O$_4$ Core/Ferrimagnetic γ-Mn$_2$O$_3$ Shell Nanoparticle System. *J. Phys.: Condens. Matter* **2011**, *23* (50), 506004.
(13)    Baaziz, W.; Pichon, B. P.; Liu, Y.; Grenèche, J.-M.; Ulhaq-Bouillet, C.; Terrier, E.; Bergeard, N.; Halté, V.; Boeglin, C.; Choueikani, F.; Toumi, M.; Mhiri, T.; Begin-Colin, S. Tuning of Synthesis Conditions by Thermal Decomposition toward Core–Shell CoxFe1–XO@CoyFe3–YO4 and CoFe2O4 Nanoparticles with Spherical and Cubic Shapes. *Chem. Mater.* **2014**, *26* (17), 5063–5073.
(14)    Meiklejohn, W. H.; Bean, C. P. New Magnetic Anisotropy. *Phys. Rev.* **1957**, *105* (3), 904–913.
(15)    Skumryev, V.; Stoyanov, S.; Zhang, Y.; Hadjipanayis, G.; Givord, D.; Nogués, J. Beating the Superparamagnetic Limit with Exchange Bias. *Nature* **2003**, *423*, 850–853.
(16)    Franceschin, G.; Gaudisson, T.; Menguy, N.; Dodrill, B. C.; Yaacoub, N.; Grenèche, J.-M.; Valenzuela, R.; Ammar, S. Exchange-Biased Fe3–xO4-CoO Granular Composites of Different Morphologies Prepared by Seed-Mediated Growth in Polyol: From Core–Shell to Multicore Embedded Structures. *Particle & Particle Systems Characterization* **2018**, *35* (8), 1800104.
(17)    Panagiotopoulos, I.; Basina, G.; Alexandrakis, V.; Devlin, E.; Hadjipanayis, G.; Colak, L.; Niarchos, D.; Tzitzios, V. Synthesis and Exchange Bias in G-Fe2O3/CoO and Reverse CoO/g-Fe2O3 Binary Nanoparticles. *The Journal of Physical Chemistry C* **2009**, *113* (33), 14609.
(18)    Baaziz, W.; Pichon, B. P.; Lefevre, C.; Ulhaq-Bouillet, C.; Greneche, J.-M.; Toumi, M.; Mhiri, T.; Begin-Colin, Sylvie. High Exchange Bias in Fe3-ΔO4@CoO Core Shell Nanoparticles Synthesized by a One-Pot Seed-Mediated Growth Method. *J. Phys. Chem. C* **2013**, *117*, 11436–11443.
(19)    Liu, X.; Pichon, B. P.; Ulhaq, C.; Lefèvre, C.; Grenèche, J.-M.; Bégin, D.; Bégin-Colin, S. Systematic Study of Exchange Coupling in Core–Shell Fe3–δO4@CoO Nanoparticles. *Chem. Mater.* **2015**, *27* (11), 4073–4081.
(20)    Liu, Y.; Liu, X.; Dolci, M.; Leuvrey, C.; Pardieu, E.; Derory, A.; Begin, D.; Begin-Colin, S.; Pichon, B. P. Investigation of the Collective Properties in Monolayers of Exchange-Biased Fe3-ΔO4@CoO Core-Shell Nanoparticles. *J. Phys. Chem. C* **2018**, *122*, 17456–17464.
(21)    Skoropata, E.; Desautels, R. D.; Chi, C.-C.; Ouyang, H.; Freeland, J. W.; van Lierop, J. Magnetism of Iron Oxide Based Core-Shell Nanoparticles from Interface Mixing with Enhanced Spin-Orbit Coupling.

*Physical Review B* **2014**, *89* (2).

(22) Broese van Groenou, A.; Bongers, P. F.; Stuyts, A. L. Magnetism, Microstructure and Crystal Chemistry of Spinel Ferrites. *Materials Science and Engineering* **1969**, *3* (6), 317–392.

(23) Sartori, K.; Cotin, G.; Bouillet, C.; Halte, V.; Begin-Colin, S.; Choueikani, F.; Pichon, B. P. Strong Interfacial Coupling through Exchange Interactions in Soft/Hard Core-Shell Nanoparticles as a Function of Cationic Distribution. *Nanoscale* **2019**, *11*, 12946–12958.

(24) Song, Q.; Zhang, Z. J. Controlled Synthesis and Magnetic Properties of Bimagnetic Spinel Ferrite $CoFe_2O_4$ and $MnFe_2O_4$ Nanocrystals with Core–Shell Architecture. *J. Am. Chem. Soc.* **2012**, *134* (24), 10182–10190.

(25) Lee, J.-H.; Jang, J.; Choi, J.; Moon, S. H.; Noh, S.; Kim, J.; Kim, J.-G.; Kim, I.-S.; Park, K. I.; Cheon, J. Exchange-Coupled Magnetic Nanoparticles for Efficient Heat Induction. *Nature Nanotechnology* **2011**, *6* (7), 418–422.

(26) Cabreira-Gomes, R.; Gomes da Silva, F.; Aquino, R.; Bonville, P.; Tourinho, F. A.; Perzynski, R.; Depeyrot, J. *Exchange Bias of MnFe2O4@gamma Fe2O3 and CoFe2O4@gamma Fe2O3 Core/Shell Nanoparticles*; 2014.

(27) Angelakeris, M.; Li, Z.-A.; Hilgendorff, M.; Simeonidis, K.; Sakellari, D.; Filippousi, M.; Tian, H.; Van Tendeloo, G.; Spasova, M.; Acet, M.; Farle, M. Enhanced Biomedical Heat-Triggered Carriers via Nanomagnetism Tuning in Ferrite-Based Nanoparticles. *Journal of Magnetism and Magnetic Materials* **2015**, *381*, 179–187.

(28) Daffé, N.; Sikora, M.; Rovezzi, M.; Bouldi, N.; Gavrilov, V.; Neveu, S.; Choueikani, F.; Ohresser, P.; Dupuis, V.; Taverna, D.; Gloter, A.; Arrio, M-A.; Sainctavit, P.; Juhin, A. Nanoscale Distribution of Magnetic Anisotropies in Bimagnetic Soft Core–Hard Shell MnFe2O4@CoFe2O4 Nanoparticles. *Advanced Materials Interfaces* **2017**, *4* (22), 1700599.

(29) Zhang, Q.; Castellanos-Rubio, I.; Munshi, R.; Orue, I.; Pelaz, B.; Gries, K. I.; Parak, W. J.; del Pino, P.; Pralle, A. Model Driven Optimization of Magnetic Anisotropy of Exchange-Coupled Core–Shell Ferrite Nanoparticles for Maximal Hysteretic Loss. *Chem. Mater.* **2015**, *27* (21), 7380–7387.

(30) Oberdick, S. D.; Abdelgawad, A.; Moya, C.; Mesbahi-Vasey, S.; Kepaptsoglou, D.; Lazarov, V. K.; Evans, R. F. L.; Meilak, D.; Skoropata, E.; van Lierop, J.; Hunt-Isaak, I.; Pan, H.; Ijiri, Y.; Krycka, K. L.; Borchers, J. A.; Majetich, S. A. Spin Canting across Core/Shell Fe3O4/MnxFe3−xO4 Nanoparticles. *Scientific Reports* **2018**, *8* (1).

(31) López-Ortega, A.; Lottini, E.; Bertoni, G.; de Julián Fernández, C.; Sangregorio, C. Topotaxial Phase Transformation in Cobalt Doped Iron Oxide Core/Shell Hard Magnetic Nanoparticles. *Chem. Mater.* **2017**, *29* (3), 1279–1289.

(32) Salazar-Alvarez, G.; Sort, J.; Uheida, A.; Muhammed, M.; Surinach, S.; Baro, M. D.; Nogues, J. Reversible Post-Synthesis Tuning of the Superparamagnetic Blocking Temperature of [Gamma]-Fe2O3 Nanoparticles by Adsorption and Desorption of Co(Ii) Ions. *Journal of Materials Chemistry* **2007**, *17* (4), 322.

(33) Juhin, A.; López-Ortega, A.; Sikora, M.; Carvallo, C.; Estrader, M.; Estradé, S.; Peiró, F.; Dolors Baró, M.; Sainctavit, P.; Glatzel, P.; Nogués, J. Direct Evidence for an Interdiffused Intermediate Layer in Bi-Magnetic Core–Shell Nanoparticles. *Nanoscale* **2014**, *6* (20), 11911–11920.

(34) Gaudisson, T.; Sayed-Hassan, R.; Yaacoub, N.; Franceschin, G.; Nowak, S.; Grenèche, J.-M.; Menguy, N.; Sainctavit, Ph.; Ammar, S. On the Exact Crystal Structure of Exchange-Biased $Fe_3O_4$–CoO Nanoaggregates Produced by Seed-Mediated Growth in Polyol. *CrystEngComm* **2016**, *18* (21), 3799–3807.

(35) Lentijo-Mozo, S.; Deiana, D.; Sogne, E.; Casu, A.; Falqui, A. Unexpected Insights about Cation-Exchange on Metal Oxide Nanoparticles and Its Effect on Their Magnetic Behavior. *Chem. Mater.* **2018**, *30* (21), 8099–8112.

(36) Salazar-Alvarez, G.; Lidbaum, H.; López-Ortega, A.; Estrader, M.; Leifer, K.; Sort, J.; Suriñach, S.; Baró, M. D.; Nogués, J. Two-, Three-, and Four-Component Magnetic Multilayer Onion Nanoparticles Based on Iron Oxides and Manganese Oxides. *J. Am. Chem. Soc.* **2011**, *133* (42), 16738–16741.

(37) Krycka, K. L.; Borchers, J. A.; Laver, M.; Salazar-Alvarez, G.; Lopez-Ortega, A.; Estrader, M.; Surinach, S.; Baro, M. D.; Sort, J.; Nogués, J. Correlating Material-Specific Layers and Magnetic

Distributions within Onion-like Fe3O4/MnO/g-Mn2O3 Core/Shell Nanoparticles. *Journal of Applied Physics* **2013**, *113* (17).
(38) Gavrilov-Isaac, V.; Neveu, S.; Dupuis, V.; Taverna, D.; Gloter, A.; Cabuil, V. Synthesis of Trimagnetic Multishell MnFe2O4@CoFe2O4@NiFe2O4 Nanoparticles. *Small* **2015**, *11* (22), 2614–2618.
(39) Sartori, K.; Choueikani, F.; Gloter, A.; Begin-Colin, S.; Taverna, D.; Pichon, B. P. Room Temperature Blocked Magnetic Nanoparticles Based on Ferrite Promoted by a Three-Step Thermal Decomposition Process. *J. Am. Chem. Soc.* **2019**, *141* (25), 9783–9787.
(40) Sartori, K.; Gailly, D.; Bouillet, C.; Grenèche, J.-M.; Dueñas-Ramirez, P.; Begin-Colin, S.; Choueikani, F.; Pichon, B. P. Increasing the Size of Fe3-ΔO4 Nanoparticles by Performing a Multistep Seed-Mediated Growth Approach. *Crystal Growth & Design* **2020**, *20* (3), 1572–1582.
(41) Cotin, G.; Kiefer, C.; Perton, F.; Boero, M.; Özdamar, B.; Bouzid, A.; Ori, G.; Massobrio, C.; Begin, D.; Pichon, B.; Mertz, D.; Begin-Colin, S. Evaluating the Critical Roles of Precursor Nature and Water Content When Tailoring Magnetic Nanoparticles for Specific Applications. *ACS Appl. Nano Mater.* **2018**, *1* (8), 4306–4316.
(42) Baaziz, W.; Pichon, B. P.; Fleutot, S.; Liu, Y.; Lefevre, C.; Greneche, J.-M.; Toumi, M.; Mhiri, T.; Begin-Colin, S. Magnetic Iron Oxide Nanoparticles: Reproducible Tuning of the Size and Nanosized-Dependent Composition, Defects, and Spin Canting. *J. Phys. Chem. C* **2014**, *118* (7), 3795–3810.
(43) Ohresser, P.; Otero, E.; Choueikani, F.; Chen, K.; Stanescu, S.; Deschamps, F.; Moreno, T.; Polack, F.; Lagarde, B.; Daguerre, J.-P.; Marteau, F.; Scheurer, F.; Joly L.; Kappler, J.-P.; Muller, B.; Bunau, O.; Sainctavit, P. DEIMOS: A Beamline Dedicated to Dichroism Measurements in the 350–2500 EV Energy Range. *Review of Scientific Instruments* **2014**, *85* (1), 013106.
(44) Daffé, N.; Choueikani, F.; Neveu, S.; Arrio, M.-A.; Juhin, A.; Ohresser, P.; Dupuis, V.; Sainctavit, P. Magnetic Anisotropies and Cationic Distribution in CoFe2O4 Nanoparticles Prepared by Co-Precipitation Route: Influence of Particle Size and Stoichiometry. *Journal of Magnetism and Magnetic Materials* **2018**, *460*, 243–252.
(45) Follath, R.; Senf, F. New Plane-Grating Monochromators for Third Generation Synchrotron Radiation Light Sources. *Nuclear Instruments and Methods in Physics Research Section A: Accelerators, Spectrometers, Detectors and Associated Equipment* **1997**, *390* (3), 388–394.
(46) Teillet, J.; Varret, F. MOSFIT Software; Université Du Maine, Le Mans, France.
(47) Ewels, P.; Sikora, T.; Serin, V.; Ewels, C. P.; Lajaunie, L. A Complete Overhaul of the Electron Energy-Loss Spectroscopy and X-Ray Absorption Spectroscopy Database: eelsdb.eu. *Microscopy and Microanalysis* **2016**, *22* (3), 717–724.
(48) Torruella, P.; Arenal, R.; de la Peña, F.; Saghi, Z.; Yedra, L.; Eljarrat, A.; López-Conesa, L.; Estrader, M.; López-Ortega, A.; Salazar-Alvarez, G.; Nogués, J.; Ducati, C.; Midgley, P. A.; Peiró, F.; Estradé S. 3D Visualization of the Iron Oxidation State in FeO/Fe3O4 Core–Shell Nanocubes from Electron Energy Loss Tomography. *Nano Lett.* **2016**, *16* (8), 5068–5073.
(49) Garnero, C.; Lepesant, M.; Garcia-Marcelot, C.; Shin, Y.; Meny, C.; Farger, P.; Warot-Fonrose, B.; Arenal, R.; Viau, G.; Soulantica, K.; Fau, P.; Poveda, P.; Lacroix, L-M.; Chaudret, B. Chemical Ordering in Bimetallic FeCo Nanoparticles: From a Direct Chemical Synthesis to Application As Efficient High-Frequency Magnetic Material. *Nano Lett.* **2019**, *19* (2), 1379–1386.
(50) Barreca, D.; Gasparotto, A.; Lebedev, O. I.; Maccato, C.; Pozza, A.; Tondello, E.; Turner, S.; Tendeloo, G. V. Controlled Vapor-Phase Synthesis of Cobalt Oxide Nanomaterials with Tuned Composition and Spatial Organization. *CrystEngComm* **2010**, *12* (7), 2185–2197.
(51) Baaziz, W.; Pichon, B. P.; Fleutot, S.; Liu, Y.; Lefevre, C.; Greneche, J.-M.; Toumi, M.; Mhiri, T.; Begin-Colin, S. Magnetic Iron Oxide Nanoparticles: Reproducible Tuning of the Size and Nanosized-Dependent Composition, Defects, and Spin Canting. *The Journal of Physical Chemistry C* **2014**, *118* (7), 3795–3810.
(52) López-Ortega, A.; Lottini, E.; Fernández, C. de J.; Sangregorio, C. Exploring the Magnetic Properties of Cobalt-Ferrite Nanoparticles for the Development of a Rare-Earth-Free Permanent Magnet. *Chemistry of Materials* **2015**, *27* (11), 4048–4056.
(53) Daou, T. J.; Grenèche, J. M.; Pourroy, G.; Buathong, S.; Derory, A.; Ulhaq-Bouillet, C.; Donnio, B.; Guillon, D.; Begin-Colin, S. Coupling Agent Effect on Magnetic Properties of Functionalized Magnetite-Based Nanoparticles. *Chem. Mater.* **2008**, *20* (18), 5869–5875.

(54) Jacintho, G. V. M.; Brolo, A. G.; Corio, P.; Suarez, P. A. Z.; Rubim, J. C. Structural Investigation of MFe$_2$O$_4$ (M = Fe, Co) Magnetic Fluids. *J. Phys. Chem. C* **2009**, *113* (18), 7684–7691.

(55) Baaziz, W.; Pichon, B. P.; Grenèche, J.-M.; Begin-Colin, S. Effect of Reaction Environment and in Situ Formation of the Precursor on the Composition and Shape of Iron Oxide Nanoparticles Synthesized by the Thermal Decomposition Method. *CrystEngComm* **2018**, No. 20, 7206.

(56) Tuček, J.; Zboril, R.; Petridis, D. Maghemite Nanoparticles by View of Mössbauer Spectroscopy. *J. Nanosci. Nanotech.* **2006**, *6* (4), 926–947.

(57) Yaacoub, N.; Mortada, H.; Nehme, Z.; Greneche, J.-M. Chemical Inhomogeneity in Iron Oxide@CoO Core–Shell Nanoparticles: A Local Probe Study Using Zero-Field and In-Field [57] Fe Mössbauer Spectrometry. *Journal of Nanoscience and Nanotechnology* **2019**, *19* (8), 5014–5019.

(58) de Bakker, P. M. A.; De Grave, E.; Vandenberghe, R. E.; Bowen, L. H. Mössbauer Study of Small-Particle Maghemite. *Hyperfine Interact* **1990**, *54* (1–4), 493–498.

(59) Daou, T. J.; Begin-Colin, S.; Grenèche, J. M.; Thomas, F.; Derory, A.; Bernhardt, P.; Legaré, P.; Pourroy, G. Phosphate Adsorption Properties of Magnetite-Based Nanoparticles. *Chem. Mater.* **2007**, *19* (18), 4494–4505.

(60) Deepak, F. L.; Bañobre-López, M.; Carbó-Argibay, E.; Cerqueira, M. F.; Piñeiro-Redondo, Y.; Rivas, J.; Thompson, C. M.; Kamali, S.; Rodríguez-Abreu, C.; Kovnir, K.; Kolen'ko, Y. V. A Systematic Study of the Structural and Magnetic Properties of Mn-, Co-, and Ni-Doped Colloidal Magnetite Nanoparticles. *J. Phys. Chem. C* **2015**, *119* (21), 11947–11957.

(61) Liu, M.; Lu, M.; Wang, L.; Xu, S.; Zhao, J.; Li, H. Mössbauer Study on the Magnetic Properties and Cation Distribution of CoFe2O4 Nanoparticles Synthesized by Hydrothermal Method. *J Mater Sci* **2016**, *51* (11), 5487–5492.

(62) Daou, T. J.; Pourroy, G.; Bégin-Colin, S.; Grenèche, J. M.; Ulhaq-Bouillet, C.; Legaré, P.; Bernhardt, P.; Leuvrey, C.; Rogez, G. Hydrothermal Synthesis of Monodisperse Magnetite Nanoparticles. *Chemistry of Materials* **2006**, *18* (18), 4399–4404.

(63) Grigorova, M.; Blythe, H. J.; Blaskov, V.; Rusanov, V.; Petkov, V.; Masheva, V.; Nihtianova, D.; Martinez, Ll. M.; Muñoz, J. S.; Mikhov, M. Magnetic Properties and Mössbauer Spectra of Nanosized CoFe2O4 Powders. *Journal of Magnetism and Magnetic Materials* **1998**, *183* (1–2), 163–172.

(64) Zhu, X.; Kalirai, S. S.; Hitchcock, A. P.; Bazylinski, D. A. What Is the Correct Fe L23 X-Ray Absorption Spectrum of Magnetite? *Journal of Electron Spectroscopy and Related Phenomena* **2015**, *199*, 19–26.

(65) Brice-Profeta, S.; Arrio, M.-A.; Tronc, E.; Menguy, N.; Letard, I.; Cartier dit Moulin, C.; Noguès, M.; Chanéac, C.; Jolivet, J.-P.; Sainctavit, Ph. Magnetic Order in - Nanoparticles: A XMCD Study. *Journal of Magnetism and Magnetic Materials* **2005**, *288*, 354–365.

(66) Torres, T. E.; Roca, A. G.; Morales, M. P.; Ibarra, A.; Marquina, C.; Ibarra, M. R.; Goya, G. F. Magnetic Properties and Energy Absorption of CoFe$_2$O$_4$ Nanoparticles for Magnetic Hyperthermia. *Journal of Physics: Conference Series* **2010**, *200* (7), 072101.

(67) Pellegrin, E.; M. Hagelstein; Doyle, S.; Moser, H. O.; Fuchs, J.; Vollath, D.; Schuppler, S.; James, M. A.; Saxena, S. S.; Niesen, L.; Rogojanu, O.; Sawatzky, G. A.; Ferrero, C.; Borowski, M.; Tjernberg, O.; Brookes, N. B. Characterization of Nanocrystalline Y-Fe2O3 with Synchrotron Radiation Techniques. *phys. stat. sol.* **1999**, *215*, 797.

(68) Li, J.; Menguy, N.; Arrio, M.-A.; Sainctavit, P.; Juhin, A.; Wang, Y.; Chen, H.; Bunau, O.; Otero, E.; Ohresser, P.; Pan, Y. Controlled Cobalt Doping in the Spinel Structure of Magnetosome Magnetite: New Evidences from Element- and Site-Specific X-Ray Magnetic Circular Dichroism Analyses. *Journal of The Royal Society Interface* **2016**, *13* (121), 20160355.

(69) Hochepied, J. F.; Sainctavit, P.; Pileni, M. P. X-Ray Absorption Spectra and X-Ray Magnetic Circular Dichroism Studies at Fe and Co L2,3 Edges of Mixed Cobalt–Zinc Ferrite Nanoparticles: Cationic Repartition, Magnetic Structure and Hysteresis Cycles. *Journal of Magnetism and Magnetic Materials* **2001**, *231* (2–3), 315–322.

(70) Haverkort, M. W. Spin and Orbital Degrees of Freedom in Transition Metal Oxides and Oxide Thin Films Studied by Soft X-Ray Absorption Spectroscopy, 2005.

(71) Byrne, J. M.; Coker, V. S.; Moise, S.; Wincott, P. L.; Vaughan, D. J.; Tuna, F.; Arenholz, E.; van der Laan, G.; Pattrick, R. A. D.; Lloyd, J. R.; Telling, N. D. Controlled Cobalt Doping in Biogenic Magnetite Nanoparticles. *Journal of The Royal Society Interface* **2013**, *10* (83), 20130134–20130134.

(72) Dolci, M.; Liu, Y.; Liu, X.; Leuvrey, C.; Derory, A.; Begin, D.; Begin-Colin, S.; Pichon, B. P. Exploring Exchange Bias Coupling in Fe3−δO4@CoO Core–Shell Nanoparticle 2D Assemblies. *Advanced Functional Materials* **2018**, *28* (26), 1706957.

(73) Pauly, M.; Pichon, B. P.; Panissod, P.; Fleutot, S.; Rodriguez, P.; Drillon, M.; Begin-Colin, Sylvie. Size Dependent Dipolar Interactions in Iron Oxide Nanoparticle Monolayer and Multilayer Langmuir-Blodgett Films. *J. Mater. Chem.* **2012**, *22*, 6343–6350.

(74) Ong, Q. K.; Wei, A.; Lin, X.-M. Exchange Bias in Fe/Fe_{3}O_{4} Core-Shell Magnetic Nanoparticles Mediated by Frozen Interfacial Spins. *Physical Review B* **2009**, *80* (13), 134418.

(75) Toulemon, D.; Pichon, B. P.; Cattoen, X.; Man, M. W. C.; Begin-Colin, Sylvie. 2D Assembly of Non-Interacting Magnetic Iron Oxide Nanoparticles via "Click" Chemistry. *Chem. Commun. (Cambridge, U. K.)* **2011**, *47*, 11954–11956.

(76) Bruvera, I. J.; Mendoza Zélis, P.; Pilar Calatayud, M.; Goya, G. F.; Sánchez, F. H. Determination of the Blocking Temperature of Magnetic Nanoparticles: The Good, the Bad, and the Ugly. *Journal of Applied Physics* **2015**, *118* (18), 184304.

(77) Shenker, H. Magnetic Anisotropy of Cobalt Ferrite (Co1.01Fe2.00O3.62) and Nickel Cobalt Ferrite (Ni0.72Fe0.20Co0.08Fe2O4). *Phys. Rev.* **1957**, *107* (5), 1246–1249.

(78) De Toro, J. A.; Marques, D. P.; Muñiz, P.; Skumryev, V.; Sort, J.; Givord, D.; Nogués, J. High Temperature Magnetic Stabilization of Cobalt Nanoparticles by an Antiferromagnetic Proximity Effect. *Phys. Rev. Lett.* **2015**, *115* (5), 057201.

(79) Nunes, W. C.; Folly, W. S. D.; Sinnecker, J. P.; Novak, M. A. Temperature Dependence of the Coercive Field in Single-Domain Particle Systems. *Phys. Rev. B* **2004**, *70* (1), 014419.

(80) Song, Q.; Zhang, Z. J. Controlled Synthesis and Magnetic Properties of Bimagnetic Spinel Ferrite CoFe$_2$O$_4$ and MnFe$_2$O$_4$ Nanocrystals with Core–Shell Architecture. *Journal of the American Chemical Society* **2012**, *134* (24), 10182–10190.

(81) Iglesias; scar; Labarta, A.; lcar; Batlle, X. Exchange Bias Phenomenology and Models of Core/Shell Nanoparticles. *Journal of Nanoscience and Nanotechnology* **2008**, *8* (6), 2761.

(82) Salazar-Alvarez, G.; Sort, J.; Uheida, A.; Muhammed, M.; Suriñach, S.; Baró, M. D.; Nogués, J. Reversible Post-Synthesis Tuning of the Superparamagnetic Blocking Temperature of γ-Fe$_2$O$_3$ Nanoparticles by Adsorption and Desorption of Co(II) Ions. *J. Mater. Chem.* **2007**, *17* (4), 322–328.

(83) Polishchuk, D.; Nedelko, N.; Solopan, S.; Ślawska-Waniewska, A.; Zamorskyi, V.; Tovstolytkin, A.; Belous, A. Profound Interfacial Effects in CoFe2O4/Fe3O4 and Fe3O4/CoFe2O4 Core/Shell Nanoparticles. *Nanoscale Research Letters* **2018**, *13* (1).

(84) López-Ortega, A.; Lottini, E.; Fernández, C. de J.; Sangregorio, C. Exploring the Magnetic Properties of Cobalt-Ferrite Nanoparticles for the Development of a Rare-Earth-Free Permanent Magnet. *Chem. Mater.* **2015**, *27* (11), 4048–4056.

(85) Torres, T. E.; Lima, E.; Mayoral, A.; Ibarra, A.; Marquina, C.; Ibarra, M. R.; Goya, G. F. Validity of the Néel-Arrhenius Model for Highly Anisotropic CoxFe3−xO4 Nanoparticles. *Journal of Applied Physics* **2015**, *118* (18), 183902.

(86) Xia, Y.; Xia, X.; Peng, H.-C. Shape-Controlled Synthesis of Colloidal Metal Nanocrystals: Thermodynamic versus Kinetic Products. *J. Am. Chem. Soc.* **2015**, *137* (25), 7947–7966.

For Table of Contents Only

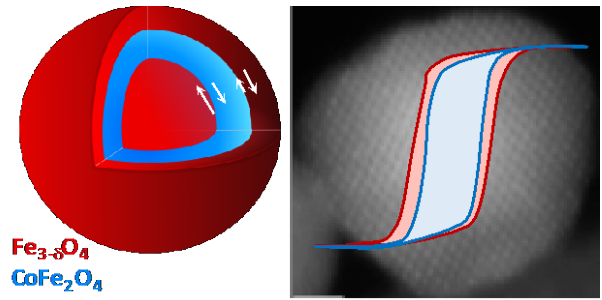